\documentclass[11pt]{article}
\usepackage{jheppub}                

\newcommand{\be}{\begin{equation}}
\newcommand{\ee}{\end{equation}}
\newcommand{\bea}{\begin{eqnarray}}
\newcommand{\eea}{\end{eqnarray}}

\newcommand{\vx}{\vec{x}}

\newcommand{\vk}{\vec{k}}

\title{On the Perturbative Stability of Quantum Field Theories in de Sitter Space}
\author[a]{D.~Boyanovsky}
\author[b]{R.~Holman}
\affiliation[a]{Department of Physics and Astronomy\\
University of Pittsburgh\\
Pittsburgh PA 15260}
\affiliation[b]{Department of Physics\\
Carnegie Mellon University\\
Pittsburgh PA 15213}
\emailAdd{boyan@pitt.edu}
\emailAdd{rh4a@andrew.cmu.edu}
\abstract{We use a field theoretic generalization of the Wigner-Weisskopf method to study the stability of the Bunch-Davies vacuum state for a massless, conformally coupled interacting test field in de Sitter space. We find that in $\lambda \phi^4$ theory the vacuum does {\em not} decay, while in non-conformally invariant models, the vacuum decays as a consequence of a vacuum wave function renormalization that depends \emph{singularly} on (conformal) time and is proportional to the spatial volume. In a particular regularization scheme the vacuum wave function renormalization is the same as in Minkowski spacetime, but in terms of the \emph{physical volume}, which leads to an interpretation of the decay. A simple example of the impact of vacuum decay upon a non-gaussian correlation is discussed. Single particle excitations also decay into two particle states, leading to particle production that hastens the exiting of modes from the de Sitter horizon resulting in the production of \emph{entangled superhorizon pairs} with a population consistent with unitary evolution. We find a non-perturbative, self-consistent ``screening'' mechanism that shuts off vacuum decay asymptotically, leading to a stationary vacuum state in a manner not unlike the approach to a fixed point in the space of states.}
\arxivnumber{}
\begin{document}
\maketitle

\section{\label{sec:intro} Introduction}

The stability of de Sitter space has always been an interesting problem to ponder, whether in the context of inflation or of the cosmological constant problem. Early studies\cite{polyakov1,IR1,IR2} revealed that de Sitter space time contains infrared instabilities and profuse particle production in interacting field theories\cite{dolgov}. Particle production in a de Sitter background has been argued to provide a ``screening'' mechanism that leads to relaxation of the cosmological constant\cite{emil,IR3,branmore} much like the production of particle-antiparticle pairs in a constant electric field. Cosmological expansion modifies the energy-uncertainty relation allowing ``virtual'' excitations to persist longer, leading to remarkable phenomena, which is stronger in de Sitter space  time\cite{woodard}.

Polyakov\cite{polyakov} has recently argued that the Bunch-Davies vacuum for interacting test scalars is unstable to vacuum decay. This process induces runaway particle production, thus rendering the in and out vacua inequivalent. Polyakov then argues that the effect of this process is to counteract the cosmological constant, not unlike   what happens in the Schwinger mechanism, leaving an non-inflationary FRW universe in its wake. While Polyakov's arguments are compelling, a firm   confirmation of this effect with a sound and systematic calculational framework is still lacking.

The arguments in ref.~\cite{polyakov} are bolstered to a certain extent by a variety of calculations of various decay rates for particle states in de Sitter space. In refs.~\cite{boyprem,boyan} as well as in ref.~\cite{moschella}, the rate for a field to decay into its own quanta was calculated, while the vacuum decay rate for a massless conformally coupled scalar with a quartic potential was computed in refs.~\cite{akhmedov,higuchi,vidal}. These latter authors use an S-matrix approach to compute the decay rate for $|0\rangle\rightarrow |\vec{k}_1,\dots,\vec{k}_4\rangle$ and argue that the lack of energy conservation allows this decay to proceed. However, it seems to us that these calculations need to be taken with a grain of salt. At the outset, the use of S-matrix amplitudes presupposes the existence of asymptotic states connected by an S-matrix that may not even exist in de Sitter space\cite{wittendeS}. More generally though, if the background is time dependent and does not turn off in the asymptotic regions, physical quantities will be time dependent and must be tracked as such.

Turning to the explicit decay rate calculations, we know that for a massless conformally coupled with a $\lambda \phi^4$ potential, we can perform the conformal rescaling to $\chi = a(\eta) \phi$, where $a(\eta)$ is the scale factor in conformal time $\eta$ to convert this to the case of a massless interacting field in Minkowski space. But the Minkowski vacuum is perturbatively stable. How can this be reconciled with the decay rate found in these afore-mentioned calculations?

In this work, we develop a different formalism, one that tracks the evolution of the quantum state directly in real time. It is the field-theoretic analog of the Wigner-Weisskopf (WW)\cite{ww} method in quantum mechanics and completely bypasses the need for the existence of asymptotic states.
In this article we introduce the method, test it in Minkowski space time and apply it to two massless conformally coupled theories in de Sitter space time: $\lambda \phi^4; \lambda \phi^3$. The latter provides the simplest realization of a non-conformally invariant theory in which de Sitter expansion leads to vacuum instability in perturbation theory.

 Our major results are that indeed we find that the $\lambda \phi^4$ vacuum is stable, as expected. The only time dependence the state acquires under time evolution is a phase corresponding to corrections to the vacuum energy. However, for non-conformally invariant interactions (e.g. cubic) the vacuum exhibits an instability to decay as a consequence of  a time dependent wavefunction renormalization  which diverges at late times, showing the infrared nature of this effect. This method allows for an unambiguous determination of the stability of the Bunch-Davies vacuum. Furthermore, it is a non-perturbative generalization of the transition amplitude one computes in perturbation theory, thus valid at late times. It is also closely related to the dynamical renormalization group method of resumming secular terms\cite{boyan,drg}, which has been used recently to deal with the secular growth of correlators in de Sitter space due to infrared effects\cite{holman}(see also refs.\cite{giddins,seery,bran}).

\section{\label{sec:WW} The Wigner-Weisskopf Method}

We start with a review of the Wigner-Weisskopf method in the case of Minkowski space time where there is
time translational invariance.

Consider a system whose Hamiltonian $H$ is given as a soluble part $H_0$ and a perturbation $H_I$: $H=H_0+H_I$. The time evolution of states in the interaction picture
of $H_0$ is given by
\be i \frac{d}{dt}|\Psi(t)\rangle_I  = H_I(t)\,|\Psi(t)\rangle_I,  \label{intpic}\ee
where the interaction Hamiltonian in the interaction picture is
\be H_I(t) = e^{iH_0\,t} H_I e^{-iH_0\,t} \label{HIoft}\ee

This has the formal solution
\be |\Psi(t)\rangle_I = U(t,t_0) |\Psi(t_0)\rangle_I \label{sol}\ee
where   the time evolution operator in the interaction picture $U(t,t_0)$ obeys \be i \frac{d}{dt}U(t,t_0)  = H_I(t)U(t,t_0)\,. \label{Ut}\ee

Now we can expand \be |\Psi(t)\rangle_I = \sum_n C_n(t) |n\rangle \label{decom}\ee where $|n\rangle$ form a complete set of orthonormal states; in the quantum field theory case these are  many-particle Fock states. From eq.(\ref{intpic}) one finds the {\em exact} equation of motion for the coefficients $C_n(t)$, namely

\be \dot{C}_n(t) = -i \sum_m C_m(t) \langle n|H_I(t)|m\rangle \,. \label{eofm}\ee

\begin{figure}[ht!]
\begin{center}
\includegraphics[height=3in,width=3in,keepaspectratio=true]{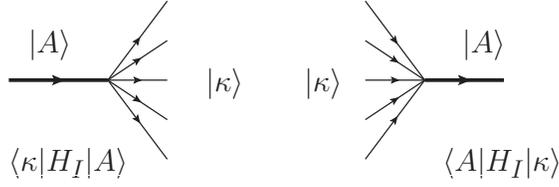}
\caption{Transitions $|A\rangle \leftrightarrow |\kappa\rangle$ in first order in $H_I$.}
\label{fig1:coupling}
\end{center}
\end{figure}

Although this equation is exact, it generates an infinite hierarchy of simultaneous equations when the Hilbert space of states spanned by $\{|n\rangle\}$ is infinite dimensional. However, this hierarchy can be truncated by considering the transition between states connected by the interaction Hamiltonian at a given order in $H_I$. Thus
consider the situation depicted in figure~\ref{fig1:coupling} where one state, $|A\rangle$, couples to a set of states $\left\{|\kappa\rangle\right\}$, which couple back only to $|A\rangle$ via $H_I$.

Under these circumstances, we have \bea \dot{C}_A(t) & = & -i \sum_{\kappa} \langle A|H_I(t)|\kappa\rangle \,C_\kappa(t)\label{CA}\\
\dot{C}_{\kappa}(t) & = & -i \, C_A(t) \langle \kappa|H_I(t) |A\rangle \label{Ckapas}\eea where the sum over $\kappa$ is over all the intermediate states coupled to $|A\rangle$ via $H_I$.

Consider the initial value problem in which at time $t=0$ the state of the system $|\Psi(t=0)\rangle = |A\rangle$ i.e. \be C_A(0)= 1,\   C_{\kappa} =0 .\label{initial}\ee  We can solve eq.(\ref{Ckapas}) and then use the solution in eq.(\ref{CA}) to find \bea  C_{\kappa}(t) & = &  -i \,\int_0^t \langle \kappa |H_I(t')|A\rangle \,C_A(t')\,dt' \label{Ckapasol}\\ \dot{C}_A(t) & = & - \int^t_0 \Sigma(t,t') \, C_A(t')\,dt' \label{intdiff} \eea where \be \Sigma(t,t') = \sum_\kappa \langle A|H_I(t)|\kappa\rangle \langle \kappa|H_I(t')|A\rangle \label{sigma} \ee This integro-differential equation  with {\em memory} yields a non-perturbative solution for the time evolution of the amplitudes and probabilities. Inserting the solution for $C_A(t)$ into eq.(\ref{Ckapasol}) one obtains the time evolution of amplitudes $C_{\kappa}(t)$ from which we can compute  the time dependent probability to populate the state $|\kappa\rangle$, $|C_\kappa(t)|^2$. This is the essence of the Weisskopf-Wigner\cite{ww} non-perturbative method ubiquitous in quantum optics\cite{qoptics}.

The hermiticity of the interaction Hamiltonian $H_I$, together with the initial conditions in eqs.(\ref{initial}) yields the unitarity condition
\be \sum_n |C_n(t)|^2 =1\,. \label{unitarity1}\ee

In general it is quite difficult to solve eq.(\ref{intdiff}) exactly, so that an approximation scheme must be developed. We do this below and then compare the approximate results to both exact results when they are available as well as other approximation schemes.

\subsection{\label{subsec:markov} Markovian approximation}

The time evolution of $C_A(t)$ determined by eq.(\ref{intdiff}) is \emph{slow} in the sense that
the time scale is determined by a weak coupling kernel $\Sigma$. This allows us to use a Markovian approximation in terms of a
consistent expansion in derivatives of $C_A$. Define \be W_0(t,t') = \int^{t'}_0 \Sigma(t,t'')dt'' \label{Wo}\ee so that \be \Sigma(t,t') = \frac{d}{dt}W_0(t,t'),\quad W_0(t,0)=0. \label{rela} \ee Integrating by parts in eq.(\ref{intdiff}) we obtain \be \int_0^t \Sigma(t,t')\,C_A(t')\, dt' = W_0(t,t)\,C_A(t) - \int_0^t W_0(t,t')\, \frac{d}{dt'}C_A(t') \,dt'. \label{marko1}\ee The second term on the right hand side is formally of \emph{fourth order} in $H_I$ and we see how a systematic approximation scheme can be developed. Setting \be W_1(t,t') = \int^{t'}_0 W_0(t,t'') dt'', \quad W_1(t,0) =0 \,\label{marko2} \ee and integrating by parts again, we find \be \int_0^t W_0(t,t')\, \frac{d}{dt'}C_A(t') \,dt' = W_1(t,t)\,\dot{C}_A(t) +\cdots \label{marko3} \ee leading to   \be \int_0^t \Sigma(t,t')\,C_A(t')\, dt' = W_0(t,t)\,C_A(t) - W_1(t,t)\,\dot{C}_A(t) +\cdots \label{histoint}\ee

This process can be implemented systematically resulting in higher order differential equations. Up to leading order in this Markovian approximation the equation eq.(\ref{intdiff}) becomes \be \dot{C}_A(t) \left[1- W_1(t,t)\right] + W_0(t,t) C_A(t) =0 \label{markovian}\ee with the result \be C_A(t) = e^{-i\int_0^t \mathcal{E}(t')dt'},\quad \mathcal{E}(t) = \frac{-i\,W_0(t,t)}{1-W_1(t,t)} \simeq -i\,W_0(t,t)\left[1+W_1(t,t)+\cdots\right] \label{solumarkov}\ee Note that in general $\mathcal{E}(t)$ is complex. In particular if the time integral in eq.(\ref{solumarkov}) is \emph{secular} in time, the real part of $\mathcal{E}$ yields a time dependent phase while the imaginary part of $\mathcal{E}$ determines the time dependent \emph{decay} since the sum rule eq.(\ref{unitarity1}) requires that the norm of the state either remains the same or diminishes. The non-secular contributions to the integral yield the overall asymptotic normalization of the state, namely the wave-function renormalization constant. The connection with the resummation scheme provided by the dynamical renormalization group will be analyzed in appendix. \ref{appsec:DRG}.

Since both $W_0,W_1 \propto H^2_I$, the leading order solution of the Markovian approximation is obtained by keeping $\mathcal{E}(t) = -i\, W_0(t,t)$. However, if $W_1(t,t)$ features secular terms that invalidate the perturbative expansion at late time then the full expression in eq.(\ref{solumarkov}) must be used.  We will see in section(\ref{sec:conjecture}) an important case where this situation emerges.

In the Markovian approximation the coefficients $C_\kappa(t)$ become \be C_{\kappa}(t)   =  -i \,\int_0^t \langle \kappa |H_I(t')|A\rangle \, e^{-i\int_0^{t'} \mathcal{E}(t'')dt''}\,dt' .\label{Ckapasolmarkov} \ee The non-perturbative nature of this solution is manifest by comparing this result to the usual perturbative transition amplitude in perturbation theory \be \mathcal{M}_{A\rightarrow \kappa} = -i\int^t_{0} \langle \kappa |H_I(t')|A\rangle \, dt'.\label{pertamp}\ee This clearly corresponds to keeping $C_A (t) =1$ in eq.(\ref{Ckapasol}), and is therefore valid only at very early times.

\subsection{\label{sec:exact} Exact Solutions}

In order to understand the domain of validity of the Markovian approximation developed in the previous subsection, we take the time in this subsection to compare its results to those obtained in a situation where time translation invariance obtains. This allows us to solve eq.(\ref{intdiff}) exactly via Laplace transform methods. Thus, suppose that $H_I$ is time independent in the Schrodinger picture. We take the states $|A\rangle$ and $\{|\kappa\rangle\}$ to be eigenstates of $H_0$, \be  H_0 |n\rangle = E_n |n\rangle \,. \label{eigen}\ee We will assume $\langle n|H_I|n\rangle =0$ by redefining the non-interacting Hamiltonian $H_0$ to include the diagonal matrix elements of the interaction; this amounts to diagonalizing the perturbation to first order in the interaction.

Using this in eqs.(\ref{CA}, \ref{Ckapas}), we find
\be C_\kappa(t) = -i \langle \kappa|H_I|A\rangle\,\int^t_0 e^{i(E_\kappa - E_A)t'}\,C_A(t') \label{Ckapaminko}\ee and \be \Sigma(t,t') = \sum_\kappa |\langle A|H_I|\kappa\rangle|^2\,e^{i(E_A-E_{\kappa})(t-t')} \equiv \int_{-\infty}^\infty d\omega'\,\rho(\omega')\, e^{i(E_A-\omega')(t-t')} \label{sigmaminko} \ee where the spectral density $\rho(\omega')$ is given by \be \rho(\omega') = \sum_\kappa |\langle A|H_I|\kappa\rangle|^2 \delta(E_\kappa-\omega')\,. \label{specdens}\ee  Introducing the Laplace variable $s$ and the Laplace transform of $C_A(t)$ as $\mathcal{C}_A(s)$, with the initial condition $C_A(t=0)=1$, we find \be \mathcal{C}_A(s)= \Bigg[s+\int_{-\infty}^\infty d\omega' ~ \frac{\rho(\omega')}{s+i(\omega'-E_A)}\Bigg]^{-1} \label{Lapla} \ee with solution \be C_A(t) = \int^{i\infty+\epsilon}_{-i\infty +\epsilon} \frac{ds}{2\pi\,i} ~\mathcal{C}_A(s)\,e^{st} \label{invlapla}\ee where the $\epsilon \rightarrow 0^+$ determines the Bromwich contour in the complex $s$-plane parallel to the imaginary axis to the right of all the singularities. Writing $s=i(\omega-i\epsilon)$ we find \be C_A(t) = \int_{-\infty}^{\infty} \frac{d\omega}{2\pi\,i}~ \frac{e^{i\omega t}}{\Bigg[\omega-i\epsilon - \int_{-\infty}^{\infty} d\omega'~\frac{\rho(\omega')}{\omega+\omega'-E_A-i\epsilon} \Bigg]  }\label{CAfin}\ee

In the free case where $\rho =0$, the pole is located at $\omega =i\epsilon \rightarrow 0$, leading to a constant $C_A$. In perturbation theory there is a complex pole very near $\omega =0$ which can be obtained directly by
expanding the integral in the denominator near $\omega =0$. We find \be  \int_{-\infty}^{\infty} d\omega'~\frac{\rho(\omega')}{\omega+\omega'-E_A-i\epsilon} \simeq -\Delta E_A - z_A\,\omega + i \,\frac{\Gamma_A}{2} \label{aproxi}\ee where \bea \Delta E_A  & = & \mathcal{P} \int_{-\infty}^{\infty} d\omega' \, \frac{\rho(\omega')}{(E_A-\omega')} \label{energyshift} \\ \Gamma_A & = & 2\pi\,\rho(E_A) \label{width} \\ z_A & = & \mathcal{P} \int_{-\infty}^{\infty} d\omega' \, \frac{\rho(\omega')}{(E_A-\omega')^2}\label{smallz}\eea and $\mathcal{P}$ stands for the principal part. The term $\Delta E_A$ is recognized as the energy shift while $\Gamma_A$ is seen to be the decay rate as found from Fermi's golden rule. The {\em long time} limit of $C_A(t)$ is determined by this complex pole near the origin leading to the asymptotic behavior \be C_A(t)\simeq \mathcal{Z}_A \, e^{-i\Delta E^r_A\,t}\,e^{-\frac{\Gamma^r_A}{2}\,t} \label{tasi}\ee where
\be \mathcal{Z}_A = \frac{1}{1+z_A}\simeq 1-z_A = \frac{\partial}{\partial E_A} \big[ E_A + \Delta E_A \big] \label{wavefunc}\ee is the wave function renormalization constant, and
\bea \Delta E^r_A & = &  \mathcal{Z}_A\,\Delta E_A \label{DeltaEr}\\ \Gamma^r_A & = &  \mathcal{Z}_A\,\Gamma_A \label{GammaAr} \eea

Now let's compare these results to those found via the Markovian approximation. With $\Sigma(t,t')$ given by eq.(\ref{sigmaminko}), to leading order in $H_I$ we find
\be \mathcal{E}(t) = -i\int^t_0 \Sigma(t,t')\,dt' =  \int_{-\infty}^{\infty} d\omega' \, \frac{\rho(\omega')}{( E_A-\omega')}\,\Bigg[ 1-e^{-i(\omega'-E_A)t} \Bigg]\label{Wominko} \ee  so that \bea \int^t_0 \mathcal{E}(t')\,dt' & = &
 t\,\int_{-\infty}^{\infty} d\omega' \, \frac{\rho(\omega')}{( E_A-\omega')}\,\Bigg[ 1-\frac{\sin(\omega'-E_A)t}{(\omega'-E_A)t} \Bigg] \nonumber \\ & - & i  \int_{-\infty}^{\infty} d\omega' \, \frac{\rho(\omega')}{( E_A-\omega')^2}\,\Bigg[ 1-\cos\big[(\omega'-E_A)t\big] \Bigg] \label{energyminko}\eea

Asymptotically as $t\rightarrow \infty$, these integrals approach:
 \be \int_{-\infty}^{\infty} d\omega' \, \frac{\rho(\omega')}{( E_A-\omega')}\,\Bigg[ 1-\frac{\sin(\omega'-E_A)\,t}{(\omega'-E_A)\,t} \Bigg] ~~\overrightarrow{t\rightarrow \infty} ~~ \mathcal{P}\int_{-\infty}^{\infty} d\omega' \, \frac{\rho(\omega')}{( E_A-\omega')} \label{realpartofE}\ee
 \be \int_{-\infty}^{\infty} d\omega' \, \frac{\rho(\omega')}{( E_A-\omega')^2}\,\Bigg[ 1-\cos\big[(\omega'-E_A)t\big] \Bigg]~~ \overrightarrow{t\rightarrow \infty} ~~\pi\,t\, \rho(E_A) + \mathcal{P} \int_{-\infty}^{\infty} d\omega' \, \frac{\rho(\omega')}{( E_A-\omega')^2} \label{imagE}\ee Using these results we find that in the late time limit,
 \be W_1(t,t)\rightarrow -\,z_A, \label{W1lt}\ee leading to the result
 \be -i \int^{t}_0 \mathcal{E}(t')\,dt' \rightarrow -i \Delta E^r_A~t - \frac{\Gamma^r_A}{2}~t -z_A \mathcal{Z}_A \label{phaseminko} \ee where $\Delta E^r_A,\Gamma_A^r,z_A$ are given by eqs. (\ref{energyshift},\ref{DeltaEr},\ref{width},\ref{GammaAr},\ref{smallz},\ref{wavefunc}). From this we read off \be C_A(t) = \mathcal{Z}_A\,e^{-i\Delta E^r_A~t}\,e^{-\frac{\Gamma^r_A}{2}~t} \label{CAmarkovres}\ee  where we approximated $e^{-z_A\,\mathcal{Z}_A} \simeq 1-z_A\,\mathcal{Z}_A =\mathcal{Z}_A$ in perturbation theory. This is in complete agreement with the asymptotic result from the exact solution eq.(\ref{tasi}).

Introducing this result into eq.(\ref{Ckapaminko}) for the coefficients $C_\kappa$ we find to leading order
 \be |C_\kappa(\infty)|^2 = \frac{|\langle \kappa|H_I|A\rangle|^2}{\Bigg[(E^R_A -E_\kappa)^2 + \frac{\Gamma^2_A}{4} \Bigg]},\quad E^R_A = E_A+\Delta E_A. \label{populationkapa}\ee This can be interpreted as follows. If $|A\rangle$ is an unstable state,  the states $|\kappa\rangle$ with $E_\kappa \sim E_A$ i.e. those nearly resonant with the state $|A\rangle$, are ``populated'' with an amplitude $\propto 1/\Gamma_A$ within a band of width $\Gamma_A$ centered at $E^R_A$. Furthermore
 \be \sum_{\kappa} |C_{\kappa}(\infty)|^2  = \int_{-\infty}^{\infty} d\omega'\, \frac{\rho(\omega')}{\Bigg[(E^R_A -\omega')^2 + \frac{\Gamma^2_A}{4} \Bigg]} \simeq 1 \label{unitaritydecay} \ee where we used eq.(\ref{width}). In this situation, unitarity entails a \emph{probability flow} towards excited states.

When $\Gamma_A =0$, i.e. $|A\rangle$ is stable,  we find \be |C_A(t)|^2 = \mathcal{Z}^2_A \simeq (1-2\,z_A). \label{nodecay}\ee Using eq.(\ref{tasi}) with $\Gamma_A=0$ in eq.(\ref{Ckapaminko}), we see that  \be |C_\kappa(t)|^2 = 2\, \frac{|\langle\kappa|H_I|A\rangle|^2}{(E^R_A-E_\kappa)^2}\Bigg[1-\cos\big[(E^R_A-E_\kappa)t \big]\Bigg]. \label{ckapanodecay}\ee In the asymptotic long time limit we find \be \sum_{\kappa \neq A}|C_\kappa(\infty)|^2 \rightarrow 2  \, \mathcal{P} \int_{-\infty}^\infty d\omega'\, \frac{\rho(\omega')}{(E^R_A-\omega')^2} = 2\,z_A \,.\label{uninodecay} \ee with $|C_A(\infty)|^2+\sum_{\kappa \neq A}|C_\kappa(\infty)|^2=1$.  This result clearly exhibits how the unitarity result in eq.(\ref{unitarity1}) is upheld in the non-decaying state case. Most of the probability remains in the initial state, although a perturbatively small amount of it, $\approx 2z$  ``flows'' to the excited states.

Our conclusion from this comparison of results is that the Markovian approximation is trustworthy at late times, which makes it the perfect tool to study vacuum stability. We can also compare its results to those of a different approximation scheme, the dynamical renormalization group\cite{boyan,drg}, which resums secular terms to extract asymptotic late time behavior. In the interests of clarity we relegate this discussion to the appendix, but the result is that the Markovian approximation also agrees with DRG results.

We now turn our attention to the field theoretic implementation of these ideas.

\section{\label{sec:WWqft} The Quantum Field Theoretic Wigner-Weisskopf Method in Minkowski Space}

In this section we apply the ideas of section \ref{sec:WW} to Minkowski space scalar field theories. We consider two separate situations. First, we take a massless scalar field with a quartic potential and compute the late time behavior of the vacuum persistence amplitude. We will then treat the case of a scalar $\varphi$ coupled to another (lighter) scalar $\chi$ via a cubic coupling $\varphi\chi^2$. This will allow us to compute both the vacuum persistence amplitude as well as the single particle amplitude, from which a decay rate can then be extracted, following the procedures described above.

\subsection{\label{subsec:quartic} Vacuum Amplitude in $\phi^4$}

We take the normal ordered interaction Hamiltonian to be \be H_I(t)= \lambda \int d^3 x :\phi^4(\vec{x},t):\,, \label{nofi4}\ee where the interaction picture field is quantized as usual in a volume $V$
\be \phi(\vec{x},t) = \frac{1}{\sqrt{V}} \sum_{\vec{k}} \frac{1}{\sqrt{2k}} \Big[ a_{\vec{k}}\, e^{-i (kt-\vec{k}\cdot\vec{x})} + a^\dagger_{\vec{k}}\, e^{i (kt-\vec{k}\cdot\vec{x})} \Big]\,.\label{field}\ee

\begin{figure}[ht!]
\begin{center}
\includegraphics[height=3in,width=3in,keepaspectratio=true]{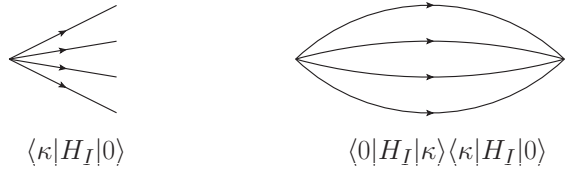}
\caption{$\varphi^4$ theory: left: transition matrix element $\langle \kappa|H_I|0\rangle$ with the four particle
state $|\kappa\rangle= |1_{\vec{k}_1};1_{\vec{k}_2};1_{\vec{k}_3};1_{\vec{k}_4}\rangle$, right: $\langle 0|H_I|\kappa\rangle \langle \kappa|H_I|0\rangle $.}
\label{fig2:fi4vacuum}
\end{center}
\end{figure}

The normal ordering cancels tadpole diagrams and makes it so that to leading order in $H_I$, the vacuum is connected only to four particle states $|\kappa\rangle= |1_{\vec{k}_1};1_{\vec{k}_2};1_{\vec{k}_3};1_{\vec{k}_4}\rangle$:

\be \langle 1_{\vec{k}1};1_{\vec{k}3};1_{\vec{k}3};1_{\vec{k}4}|H_I(t)|0\rangle = \lambda\,\frac{(2\pi)^3}{V^2}\,\delta^{(3)}\Big( \sum_{i=1}^4 \vec{k}_i\Big)\,\frac{e^{i(k_1+k_2+k_3+k_4)t}}{4\left(k_1 k_2 k_3 k_4 \right)^\frac{1}{2}}; \label{mtxele}\ee
we have subtracted the zero point energy of the vacuum so that $H_0|0\rangle =0$.  This matrix element is depicted by the left hand side in figure \ref{fig2:fi4vacuum}.  The kernel $\Sigma(t,t')$ of eq.(\ref{sigmaminko}) is given by
\be \Sigma(t,t') = \lambda^2\,\sum_{\vec{k}_1}\cdots\sum_{\vec{k}_4}\frac{(2\pi)^3\,V\,\delta^{(3)}\Big( \sum_{i=1}^4 \vec{k}_i\Big)}{V^4\,16\,k_1 k_2 k_3 k_4}\,e^{-i(k_1+k_2+k_3+k_4)(t-t')} = \int_{-\infty}^\infty d\omega'\,\rho(\omega')\,e^{-i\omega'(t-t')}\label{sigfi4mink}\ee where
\be \rho(\omega') = \lambda^2\, V\, \int \frac{d^3k_1}{(2\pi)^3} \int \frac{d^3k_2}{(2\pi)^3} \int \frac{d^3k_3}{(2\pi)^3} \frac{\delta(k_1+k_2+k_3+|\vec{k}_1+\vec{k}_2+\vec{k}_3|-\omega')}{16\ k_1 k_2 k_3 |\vec{k}_1+\vec{k}_2+\vec{k}_3|}. \label{rhofi4}\ee This describes the \emph{three loop} diagram on the right hand side of figure \ref{fig2:fi4vacuum}.

From the results of the previous section we find
\be C_0(t) = e^{-z_0}\,e^{-i\Delta E_0\,t} \label{c0finali}\,,\ee where
\bea \Delta E_0 &=& - \lambda^2\,V\,\int  \prod_{i=1}^3 \frac{d^3k_i}{(2\pi)^3}  \frac{1}{16 \left(k_1 k_2 k_3 |\vec{k}_1+\vec{k}_2+\vec{k}_3|\right)}\frac{1}{\left(\sum_{i=1}^3 k_i+ |\vec{k}_1+\vec{k}_2+\vec{k}_3|\right)}\label{deltaE0fi4}\\
z_0 &=& \lambda^2\,V\, \int  \prod_{i=1}^3 \frac{d^3k_i}{(2\pi)^3} \frac{1}{16 \left(k_1 k_2 k_3 |\vec{k}_1+\vec{k}_2+\vec{k}_3|\right)}\frac{1}{\left(\sum_{i=1}^3 k_i+ |\vec{k}_1+\vec{k}_2+\vec{k}_3|\right)^2}\label{z0fi4}\eea

We also find that the vacuum is stable \be \Gamma_0 = 2\pi \rho(\omega'=0) = 0 \label{gamma0fi4}\ee as expected.

$\Delta E_0$ is recognized as the leading order $\mathcal{O}(\lambda^2)$ correction to the vacuum energy in the normal ordered theory. The volume factor just reflects the extensivity of the vacuum energy.

As we will compare the results obtained in Minkowski space-time with the case of de  Sitter cosmology in the next section, it will prove illuminating to understand eqs.(\ref{deltaE0fi4}, \ref{gamma0fi4}, \ref{z0fi4}) in more detail.
According to the adiabatic theorem the exact ground state $|\widetilde{0 }\rangle$ is obtained from the non-interacting vacuum $|0\rangle$ by adiabatically switching on the perturbation $H_I \rightarrow e^{-\epsilon |t|}\,H_I$, with $\epsilon \rightarrow 0^+$
\be |\widetilde{0 }\rangle = U_\epsilon(0,-\infty)|0\rangle \label{adiab}\ee where $U_\epsilon(t,t_0)$ is the time evolution in operator in the interaction picture with adiabatic switching-on of the interaction Hamiltonian. To $\mathcal{O}(\lambda)$ we find
\be |\widetilde{0 }\rangle =  {|0\rangle}-i  \int^0_{-\infty}\sum_\kappa |\kappa\rangle \langle\kappa|H_I(t)|0\rangle\,e^{-\epsilon |t|} dt \label{dressed}\ee where $|\kappa\rangle$ are the four particle states
and $\langle \kappa|H_I(t)|0\rangle$ is given by eq.(\ref{mtxele}). With $H_I(t) = e^{iH_0t}H_I e^{-iH_0t}$ the time integral can be done straightforwardly leading to
\be \int^0_{-\infty}  \langle\kappa|H_I(t)|0\rangle\,e^{-\epsilon |t|}\, dt = -i\lambda \, \frac{(2\pi)^3}{V^2}\,\delta^{(3)}\Big( \sum_{i=1}^4 \vec{k}_i\Big)\,\frac{1}{4\left(k_1 k_2 k_3 k_4 \right)^\frac{1}{2}}
\frac{1}{\left(k_1+k_2+k_3+k_4+i\epsilon\right)} \label{adiainteg}\ee

It is clear that the transition amplitude $\langle \kappa|U_\epsilon(0,-\infty)|0\rangle$ has nothing to do with the decay of the vacuum state, which is stable in Minkowski spacetime, but rather with the ``dressing'' of the vacuum by virtual excitations; the true vacuum state is a superposition of many particle Fock states. Comparing eq.(\ref{adiainteg}) with eq.(\ref{z0fi4}) it is clear that
\be z_0 = \sum_\kappa |\langle \kappa |\widetilde{0}\rangle|^2 \label{overli}\ee is given by eq.(\ref{z0fi4}) and is a manifestation of unitarity. Furthermore the exact energy of the adiabatically constructed ground state is given by \be E = E_0+\Delta E_0
= \frac{\langle \widetilde{0}|(H_0+H_I)|\widetilde{0}\rangle}{\langle \widetilde{0}|\widetilde{0}\rangle }\ee which up to $\mathcal{O}(\lambda^2)$ is given by eq.(\ref{deltaE0fi4}). This example clarifies that the Weisskopf-Wigner method provides a non-perturbative alternative to the adiabatic theorem of Gell-Mann and Low\cite{gellmannlow}.

\subsection{\label{subsec:cubic} Vacuum amplitude and $1$-particle decay in $\varphi \chi^2$}

The cubic coupling between two fields gives us the opportunity to compute a non-trivial decay rate using the WW techniques. We first compute the vacuum amplitude as we did in the quartic case, though in this situation instead of dealing with a three-loop diagram, the corrections come in at two loops.

\begin{figure}[ht!]
\begin{center}
\includegraphics[height=3in,width=3in,keepaspectratio=true]{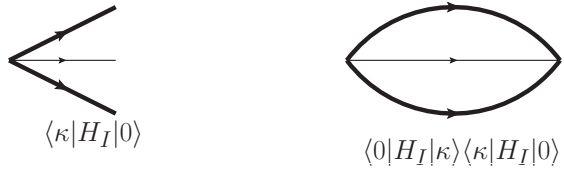}
\caption{$\varphi \,\chi^2$ vacuum to vacuum amplitude. Thin lines correspond to $\varphi$, thick lines to $\chi$. }
\label{fig3:fi3vaccum}
\end{center}
\end{figure}

Thus, consider a massive field $\varphi$ coupled to a massless field $\chi$ with the interaction Hamiltonian being
\be H_I = \lambda \int d^3x\ \varphi(\vec{x})\,\chi^2(\vec{x}). \label{cubic}\ee The leading order vacuum to vacuum matrix elements are shown in figure \ref{fig3:fi3vaccum} and we find \be \Sigma_0(t,t') = \int_{-\infty}^\infty d\omega'\,\rho_0(\omega')\,e^{-i\omega'(t-t')} \label{Sigma0fi3}\ee with
\be \rho_0(\omega') = V  {\lambda^2}  \int \frac{d^3p}{(2\pi)^3}\int \frac{d^3k }{(2\pi)^3} \frac{\delta(\omega'-E_p-k -|\vec{k}+\vec{p}|)}{2E_p\,2k \,2|\vec{k}+\vec{p}|},\quad E_p=\sqrt{p^2+m^2_\varphi}. \label{rho0fi3}\ee We also have that
\be W_0(t,t) = -i  \int_{-\infty}^\infty d\omega'\,\frac{\rho_0(\omega')}{\omega'}\,\left(1-e^{-i\omega't}\right).\label{Wofi3}\ee Taking the long time limit of the leading order contribution, we find, using eqs.(\ref{realpartofE},\ref{imagE}):
\be -i \int^t_0 \mathcal{E}(t') dt'\rightarrow -i\Delta E_0\,t -z_0 \label{asifi3}\ee where
\be \Delta E_0 = -\mathcal{P}\int_{-\infty}^\infty d\omega' \frac{\rho_0(\omega')}{ \omega'},\quad z_0  =  \mathcal{P}\int_{-\infty}^\infty d\omega' \frac{\rho_0(\omega')}{\omega^{'\,2}}; \label{fi3phase}\ee note that $\Delta E_0$ and $z_0$ are quadratically and linearly divergent in the ultraviolet respectively. This observation will be important when we analyze the vacuum to vacuum amplitude in de Sitter space-time.
The late time behavior of the vacuum persistence amplitude is found to be
\be C_0(t) = \mathcal{Z}_0\, e^{-i\Delta E_0\,t},\quad \mathcal{Z}_0 = e^{-z_0} \simeq 1-z_0. \label{C0fi3}\ee

The diagrams that are summed by this result can be clearly interpreted: they are those that feature the leading dependence on the volume $V$ at each order in $\lambda^2$  . For example at order $\lambda^4$ there are two types of contributions, two disconnected diagrams of the type depicted in figure \ref{fig3:fi3vaccum} with a contribution $\propto (\lambda^2 V)^2$, and another disconnected diagram in which one of the internal lines features a one-loop self energy correction, with a contribution $\propto \lambda^4 \, V$ which is subleading in terms of the volume factors and represents the $\mathcal{O}(\lambda^4)$ correction to the vacuum energy.

\begin{figure}[htbp!]
\begin{center}
\includegraphics[height=3in,width=3in,keepaspectratio=true]{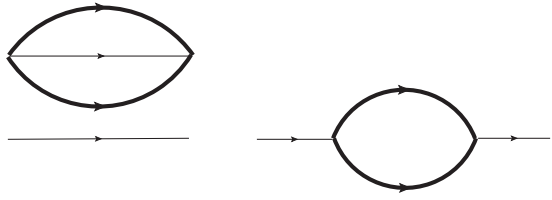}
\caption{$\varphi\,\chi^2$ one $\varphi$ particle amplitude. Thin lines correspond to $\varphi$, thick lines to $\chi$.}
\label{fig4:oneparticle}
\end{center}
\end{figure}

Next we turn our attention to the behavior of the single particle amplitudes $ C_{1^\varphi_p}(t)$. To leading order in $H_I$, the one $\varphi$ state is connected to states with two-$\chi$ quanta. There are two contributions to $\Sigma(t,t')$ in this case: a disconnected one, describing a change in the vacuum state as well as a connected one which incorporates the contribution of the decay process to the self-energy of the $\varphi$ field (figure \ref{fig4:oneparticle}).

The two-$\chi$ states for each diagram are different. For the vacuum diagram, we have  $|\kappa\rangle=|1^\varphi_{\vec{p}},1^\varphi_{\vec{k}_1},1^\chi_{\vec{k}_2},1^\chi_{\vec{k}_3}\rangle$ with $\vec{k}_1+\vec{k}_2+\vec{k}_3 =\vec{0}$, while the connected diagram requires $|\kappa\rangle=|1^\chi_{\vec{k}_1},1^\chi_{\vec{k}_2}\rangle$ with $\vec{p} = \vec{k}_1+\vec{k}_2$.

It is straightforward to find in this case \be \Sigma(t,t') = \Sigma_0(t,t')+\Sigma_1(\vec{p},t,t') \label{sig1pfi3}\ee where $\Sigma_0(t,t')$ is given by eqs.(\ref{Sigma0fi3},\ref{rho0fi3}) and
\be \Sigma_1(\vec{p},t,t') = \int^\infty_{-\infty} \rho_1(p,\omega') \, e^{-i(\omega'-E_p)(t-t')} \label{sigma1fi3}\ee with
\be \rho_1(p,\omega') = \lambda^2 \int \frac{d^3k}{(2\pi)^3}\,\frac{\delta(\omega'-k-|\vec{p}-\vec{k}|)}{2E_p\,2k\,2|\vec{p}-\vec{k}|} \label{rho1fi3}\ee Following the same steps as in the previous examples we find
 \be -i \int^t_0 \mathcal{E}(t') dt'\rightarrow -i\Big[\Delta E_0+\Delta E_1(p)\Big]\,t-\frac{\Gamma_\varphi}{2}t  -\big[z_0+z_1(p)\big] \label{asi1parfi3}\ee
 where $\Delta E_0$ and $z_0$ are given by eq.(\ref{fi3phase})  and
 \be \Delta E_1(p) = \mathcal{P}\int_{-\infty}^\infty d\omega' \frac{\rho_1(p,\omega')}{E_p- \omega'}, \quad z_1(p) =  \mathcal{P}\int_{-\infty}^\infty d\omega' \frac{\rho_1(\omega')}{(E_p-\omega')^2}. \label{1pfi3phase} \ee
 The decay rate for $\varphi \rightarrow 2\chi$ is found to be
 \be \Gamma_\varphi = 2\pi \rho_1(p,E_p) = 2\pi \lambda^2 \int \frac{d^3k}{(2\pi)^3}\,\frac{\delta(E_p-k-|\vec{p}-\vec{k}|)}{2E_p\,2k\,2|\vec{p}-\vec{k}|}, \label{gamma1pfi3}\ee leading to
 \be C_{1^\varphi_p}(t) = \mathcal{Z}_0 \mathcal{Z}_{1}(p)\,e^{-i\Delta E_0 t}\,e^{-i\Delta E_1(p)t}\,e^{-\frac{\Gamma_\varphi}{2}t} \label{C1fi3}\ee

The interpretation of this result is clear: $\Delta E_0 + \Delta E_1(p)$ is the \emph{total} energy shift of the one particle state with respect to the unperturbed vacuum state, whereas $\Delta E_1(p)$ is the energy shift with respect to the ``dressed'' vacuum state.  Similarly, the \emph{total} wave function renormalization $\mathcal{Z}_0 \mathcal{Z}_1(p)$ is the overlap between the bare single particle state obtained from the bare vacuum by applying a creation operator and the full ``dressed'' single particle state.

 The phase $\Delta E_0 \, t$ is the vacuum energy and is common to all many particle states. It can be renormalized away by introducing a counterterm Hamiltonian proportional to the identity operator
\be H_{ct} = - \Delta E_0. \label{HctMinko} \ee Then each many particle state features a term in the equation of motion for their amplitudes $C_n$ given by  $i \Delta E_0 \, C_n$, which exactly cancels the phase from the vacuum energy (disconnected diagram). Furthermore, renormalizing the vacuum state and defining single particle states with respect to the full ``dressed'' vacuum state, can be achieved by defining the \emph{renormalized} amplitudes
 \be \widetilde{C}_n(t) = C_n(t) \mathcal{Z}_0^{-1} \label{renamps}\ee  leading to
 \be \widetilde{C}_{1^\varphi_p}(t) =   \mathcal{Z}_{1}(p) \,e^{-i\Delta E_1(p)t}\,e^{-\frac{\Gamma_\varphi}{2}t} \label{C1fi3ren}\ee

 This renormalization procedure  cancels the contribution of the disconnected diagram which multiplies all many particle amplitudes by $\mathcal{Z}_0 \,e^{-i\Delta E_0\,t}$. This is  familiar from renormalized perturbation theory; the vacuum diagrams cancel between numerators and denominators in expectation values or correlation functions, since these correspond to the renormalization of the vacuum state. However, these disconnected graphs are present in the time evolution of the quantum states themselves.The redefinition of particle states by dividing the ``bare'' amplitudes by the amplitude of finding the bare state in the ``dressed'' state is an important ingredient in the adiabatic hypothesis of Gell-Mann and Low\cite{gellmannlow}.

\subsubsection{\label{subsubsec:partprod} Particle decay and particle production}

The decay of a $\varphi$ particle of momentum $\vec{p}$ results in the production of  two $\chi$ particles with $\vec{k}_1,\vec{k}_2$. The final result for the amplitude of the state with a pair of $\chi$ particles can be
understood from a simple argument which will be useful in the discussion of particle production in de Sitter spacetime. From eq.(\ref{Ckapas}) the amplitude of the two $\chi$ particle state is given by
 \be \widetilde{C}_{2\,\chi}(t) = -i \langle 1^\chi_{\vec{k}_1},1^\chi_{\vec{k}_2}|H_I|1^\varphi_{\vec{p}}\rangle \,
 \int^t_0\,e^{-i(E_p-k-|\vec{p}-\vec{k}|)t'}\,\widetilde{C}_{1^\varphi_p}(t')\,dt'. \label{pairamp}\ee Taking the amplitude $C_{1^\varphi_p}(t')\simeq 1$ during the decay time scale $t' \lesssim 1/\Gamma_\varphi$, the integrand in (\ref{pairamp}) is $\simeq 1$ for the resonant wave-vectors $ k+|\vec{p}-\vec{k}| \sim E_p$ and the integral yields a secular term that grows linearly with time, while for wave-vectors well outside the resonant band, the
 rapid phase averages the integral out. Since the amplitude $\widetilde{C}_{1^\varphi_p}(t') \sim 0$ for $t'>>1/\Gamma_\varphi$, the resonant wave-vectors for which the phase nearly vanishes lead to
 \be \widetilde{C}_{2\,\chi}(t>>1/\Gamma_\varphi) \propto 1/\Gamma_\varphi. \label{secu}\ee Thus ``resonant pairs'' are produced linearly in time within a time $\sim 1/\Gamma_\varphi$ during which the amplitude $C_{1^\varphi_p}$ is nearly  constant. This argument confirms the general result eq.(\ref{populationkapa}), and will be useful in the interpretation of particle production from decay in de Sitter spacetime.

\subsubsection{\label{subsubsec:wavefunc} On the wave function renormalization of the vacuum state:}

 In anticipation of a study of the time evolution of quantum state in de Sitter spacetime we analyze in detail the time evolution of the vacuum state in the case in which all fields $\varphi,\chi$ are \emph{massless}, because we want to use this study in Minkowski spacetime as a guideline for interpretation of results in de Sitter cosmology. In this case $\Sigma_0(t,t')$ can be calculated exactly by carrying out the \emph{three body phase space} momentum integrals in eq.(\ref{Sigma0fi3},\ref{rho0fi3}), and we find
 \be \Sigma_0(t,t') = \frac{i\,\lambda^2 V}{(4\pi)^4\,(t-t'-i\epsilon)^3},\quad \epsilon \rightarrow 0^+ \label{sig0tfi3}\ee where $\epsilon$ provides a convergence factor that regularizes the ultraviolet behavior of the integrals. From this it follows that
 \be -i\int_0^t \mathcal{E}(t')\, dt' = -i\frac{\lambda^2 V}{2(4\pi)^4}\left[-\frac{t}{\epsilon^2}-\frac{i}{\epsilon}+\frac{1}{t}\right] \label{phasemasslessfi3}\ee from which we can extract in the long time limit
 \be \Delta E_0 = -\frac{\lambda^2V}{2(4\pi)^4\,\epsilon^2},\quad z_0 = \frac{\lambda^2 V}{2(4\pi)^4\,\epsilon}. \label{fasefi3eps}\ee This clearly  displays the quadratic ($1/\epsilon^2$) and linear ($1/\epsilon$) ultraviolet divergences of the energy shift and wave function renormalization respectively.

 We also find
 \be W_0(t,t') = i \,\frac{\lambda^2 \,V}{2(4\pi)^4} \left[\frac{1}{(t-t'-i\epsilon)^2} - \frac{1}{(t-i\epsilon)^2} \right] \label{Wominkfi3m0} \ee and
 \be W_1(t,t) = -\frac{\lambda^2 \,V}{2(4\pi)^4\,\epsilon} \left[1+ i\,\frac{\epsilon}{t}\right] = -z_0 +\mathcal{O}(1/t), \label{W1minkfi3m0}\ee this last result confirming the asymptotic behavior of  eq.(\ref{W1lt}).  A similar calculation shows that the wave-function renormalization for the single particle state, $z_1(p)$, is ultraviolet finite.

It should be clear from the above discussion, and the comparison with the exact solution using the Laplace transform, that the vacuum amplitude $C_0(t)$ provides a resummation of the vacuum diagrams of the type depicted in figures (\ref{fig2:fi4vacuum},\ref{fig3:fi3vaccum}), whereas the amplitude for the single particle state $C_1(t)$ provides a Dyson-like resummation of the self-energy diagrams depicted in figure \ref{fig4:oneparticle}.

This discussion and the comparison to known results in Minkowski space time explicitly clarifies that the WW method provides a resummation of the perturbative series akin to the Dyson resummation of self-energies insertion but directly in real time. For the vacuum case it sums (to lowest order) the two ($\phi^3$) or three ($\phi^4$) loop diagrams for the vacuum energy, and  for single particle states it resums (to lowest order) the one-loop self energy diagrams (for $\phi^3$) along with the vacuum diagrams. The loop corrections are given
by the usual momentum integrals.


Although the $\phi^3$ theory is unbounded below and therefore has an intrinsic instability, such instability is \emph{not} manifest either in the vacuum wave function renormalization or in single particle states for the self-interacting case ($\chi=\phi$) since kinematics prevents $1\rightarrow 2$ processes.

We will resort to these results when we try to interpret the various contributions to the evolution of the vacuum state in de Sitter spacetime.

\section{\label{sec:WWdeSqft} The Quantum Field Theoretic Wigner-Weisskopf Method in de Sitter Space}

The previous section showed quite clearly that the solution to the WW equations given by the Markovian approximation reproduces the results of standard flat space perturbation theory \emph{and} and provides   non-perturbative control of the late time behavior of the vacuum and many particle amplitudes $C_n(t)$. Having established the reliability of the non-perturbative result in Minkowski spacetime, we now turn to studying the time evolution of states in de Sitter spacetimes.

We will consider the same situations as described in the previous section, i.e. first that of a quartically self-coupled field and then a cubically coupled one. The new features to be added to the discussion are first, that we will consider only  massless, conformally coupled fields (i.e. $\xi=1\slash 6$), and second, in the cubic case we allow for the possibility that $\chi=\varphi$. In Minkowski spacetime this last condition would preclude any decay phenomena from occurring, but as we shall see below, the lack of a global time-like Killing vector in de Sitter spacetime allows energy non-conserving processes to occur.

We consider a spatially flat Friedmann-Robertson-Walker (FRW)
cosmological spacetime with scale factor $a(t)$. In comoving
coordinates, the action is given by
\begin{equation}
S= \int d^3x \; dt \;  a^3(t) \Bigg\{
\frac{1}{2}{\dot{\phi}^2}-\frac{(\nabla
\phi)^2}{2a^2}-\frac{1}{2}\Big(M^2+\xi \; \mathcal{R}\Big)\phi^2
- g \;  \phi^{\,p}  \Bigg\},\quad p=3,4 \label{lagrads}
\end{equation}
with \be \mathcal{R} = 6 \left(
\frac{\ddot{a}}{a}+\frac{\dot{a}^2}{a^2}\right) \ee being the Ricci
scalar,   $\xi=
0$ corresponds to minimal coupling while $\xi=1/6$ corresponds to
conformal coupling.

Specializing now to the de Sitter case with $a(t) = e^{H t}$, it is convenient to pass to conformal time $\eta = -e^{-Ht}/H$ with $d\eta =
dt/a(t)$ and introduce a conformal rescaling of the fields
\begin{equation}
a(t)\phi(\vx,t) = \chi(\vx,\eta).\label{rescale}
\end{equation}
The action becomes (after discarding surface terms that will not
change the equations of motion) \be S  =
\frac12 \int d^3x \; d\eta  \; \Bigg\{\frac12\left[
{\chi'}^2-(\nabla \chi)^2-\mathcal{M}^2_{\chi}(\eta) \; \chi^2  \right] -\lambda \big[C (\eta)\big]^{(4-p)} \;  \chi^{\,p}   \Bigg\} \; , \label{rescalagds}\ee
with primes denoting derivatives with respect to
conformal time $\eta$ and \be \mathcal{M}^2_{\chi}(\eta) =
\Big(M^2+\xi \mathcal{R}\Big)
C^2(\eta)-\frac{C''(\eta)}{C(\eta)}  \; , \label{massds}
\ee
where for de Sitter spacetime \be C(\eta)= a(t(\eta))= -\frac{1}{H\eta}. \label{scalefactords}\ee  In this case the effective time dependent mass is given by
\be
\mathcal{M}^2_{\chi}(\eta)  = \Big[\frac{M^2}{H^2}+12\Big(\xi -
\frac{1}{6}\Big)\Big]\frac{1}{\eta^2}   \; . \label{massds2}
\ee
The Heisenberg equations of motion for the spatial Fourier modes
of wavevector $k$  of the fields in the non-interacting ($g=0$)
theory are given by
\be  \chi''_{\vk}(\eta)+
\Big[k^2-\frac{1}{\eta^2}\Big(\nu^2 -\frac{1}{4} \Big)
\Big]\chi_{\vk}(\eta)  =   0   \label{dsmodes}  \; ,
\ee
\noindent where
\be
\nu^2_{\chi}  =  \frac{9}{4}-\Big( \frac{M^2}{H^2}+12\,\xi
\Big)   \label{nu}\ee
Two linearly independent solutions are given by
\bea
g_{\nu}(k;\eta) & = & \frac{1}{2}\; i^{-\nu-\frac{1}{2}}
\sqrt{\pi \eta}\,H^{(2)}_\nu(k\eta)\label{gnu}\\
f_{\nu}(k;\eta) & = & \frac{1}{2}\; i^{\nu+\frac{1}{2}}
\sqrt{\pi \eta}\,H^{(1)}_\nu(k\eta)= [g_{\nu}(k;\eta)]^*\label{fnu}  \; ,
\eea
 \noindent where $H^{(1,2)}_\nu(z)$ are Hankel functions. Expanding the field operator in this basis
\be \chi(\vec{x},\eta) = \frac{1}{\sqrt{V}}\sum_{\vec{k}} \Big[a_{\vec{k}}\,g_\nu(k;\eta)\,e^{i\vec{k}\cdot\vec{x}}+ a^\dagger_{\vec{k}}\,\,f_\nu(k;\eta)\,e^{-i\vec{k}\cdot\vec{x}}\Big]\,. \label{quantumfieldds1}\ee
The Bunch-Davies vacuum is defined such that \be a_{\vec{k}}|0\rangle = 0 \,,\label{dsvac}\ee and the Fock space states are obtained in the usual manner, i.e. by applying creation operators $a_{\vec{k}}^{\dagger}$ to the vacuum.

 In the Schroedinger picture the quantum states $|\Psi(\eta)\rangle$ obey
 \be i\frac{d}{d\eta}|\Psi(\eta)\rangle = H(\eta) \, |\Psi(\eta)\rangle \label{Spic}\ee where in an expanding cosmology the Hamiltonian $H(\eta)$ is generally a function of $\eta$ (unlike in Minkowski space-time where it is a constant by energy conservation). Introducing the time evolution operator $U(\eta,\eta_0)$ obeying
 \be i\frac{d}{d\eta} U(\eta,\eta_0) = H(\eta)\,U(\eta,\eta_0) , \quad U(\eta_0,\eta_0) = 1, \label{Uds}\ee the solution of the Schroedinger equation is $|\Psi(\eta)\rangle = U(\eta,\eta_0)\,|\Psi(\eta_0)\rangle $. Writing the Hamiltonian as $H(\eta) = H_0(\eta) + H_{I}(\eta)$ with $H_0(\eta)$ the non-interacting Hamiltonian, and introducing the time evolution operator of the free theory $U_0(\eta,\eta_0)$ satisfying
 \be i\frac{d}{d\eta} U_0(\eta,\eta_0) = H_0(\eta)\, U_0(\eta,\eta_0), \quad i\frac{d}{d\eta} U^{-1}_0(\eta,\eta_0) = - U^{-1}_0(\eta,\eta_0)\, H_0(\eta) , \quad U_0(\eta_0,\eta_0) =1, \label{U0ds}\ee the interaction picture states are defined as
 \be |\Psi(\eta)\rangle_I = U^{-1}_0(\eta,\eta_0) |\Psi(\eta)\rangle. \label{ipds}\ee These obey
 \be  \frac{d}{d\eta}|\Psi(\eta)\rangle_I = -i H_I(\eta) |\Psi(\eta)\rangle_I, \quad H_I(\eta) = U^{-1}_0(\eta,\eta_0) H_{I}(\eta) U_0(\eta,\eta_0) \label{HIdsdef}\ee  where the normal-ordered interaction Hamiltonian in the interaction picture is
 \be H_I(\eta) = \frac{\lambda}{[-H\eta]^{(4-p)}}\int  :[\chi(\vec{x},\eta)]^p: d^3 x \label{HIds}\ee where $\chi$ is the free field Heisenberg field operator in eq.(\ref{quantumfieldds1}). We now expand  the state $|\Psi(\eta)\rangle_I$ in Bunch-Davies Fock states $|n\rangle$  $ |\Psi(\eta)\rangle_I = \sum_n C_n(\eta) |n\rangle$.

We can now implement the Weisskopf-Wigner program directly by replacing $t \rightarrow \eta$ in the results of the previous sections, with the proviso that $t\rightarrow \infty \Rightarrow  \eta \rightarrow 0^-$, and the initial conditions on the states are given at some given initial time $\eta_0$. Eventually we will take $\eta_0 \rightarrow -\infty$. Equations (\ref{Ckapasol},\ref{intdiff}) now become \bea  C_{\kappa}(\eta) & = &  -i \,\int_{\eta_0}^{\eta} \langle \kappa |H_I(\eta')|A\rangle \,C_A(\eta')\,d\eta' \label{Ckapasolds}\\ \dot{C}_A(\eta) & = & - \int^{\eta}_{\eta_0} \Sigma(\eta,\eta') \, C_A(\eta')\,d\eta' \label{intdiffds} \eea where \be \Sigma(\eta,\eta') = \sum_\kappa \langle A|H_I(\eta)|\kappa\rangle \langle \kappa|H_I(\eta')|A\rangle \label{sigmads} \ee and the intial conditions $C_A(\eta_0)=1,\ C_{\kappa \neq A}(\eta_0) =0$.

 In the Markovian approximation to leading order in the interaction
 \be C_A(\eta) = e^{-\int^{\eta}_{\eta_0}W_0(\eta',\eta')\, d\eta'}  , \quad W_0(\eta',\eta')=\int_{\eta_0}^{\eta'} \Sigma(\eta',\eta^{''}) d\eta^{''}\,. \label{dssolu} \ee

We now specialize to the case described at the beginning of this section, namely $M^2=0,\ \xi =1/6,\ \nu=1/2$, in which case canonical quantization of the scalar field in a finite comoving volume $V$ leads to
 \be \chi(\vec{x},\eta) = \frac{1}{\sqrt{V}}\sum_{\vec{k}} \frac{1}{\sqrt{2k}}\Big[a_{\vec{k}}\,e^{-ik\eta}\,e^{i\vec{k}\cdot\vec{x}}+ a^\dagger_{\vec{k}}\,e^{ ik\eta}\,e^{-i\vec{k}\cdot\vec{x}}\Big]\,. \label{quantumfieldds}\ee

 \subsection{\label{subsec:quarticdeS} De Sitter Vacuum Amplitude in $\phi^4$}

A massless, conformally coupled quartically self-coupled scalar theory is conformally (or more precisely Weyl) invariant and to lowest order (i.e. neglecting higher order radiative corrections that break conformal invariance via renormalization effects), correlation functions are independent of the metric.

 The vacuum to vacuum amplitude is described once again by the diagram in figure (\ref{fig2:fi4vacuum}) leading to
 \be \Sigma_0(\eta,\eta') = \lambda^2   \, V\, \int \prod_{i=1}^3 \frac{d^3k_i}{(2\pi)^3} \frac{e^{-i(\sum_{i=1}^3 k_i+|\vec{k}_1+\vec{k}_2+\vec{k}_3|)(\eta-\eta')}}{16 k_1 k_2 k_3 |\vec{k}_1+\vec{k}_2+\vec{k}_3|} \label{sigmafi4ds}\ee
 \be W_0(\eta',\eta') = -i \lambda^2   \, V\, \int \prod_{i=1}^3 \frac{d^3k_i}{(2\pi)^3} \frac{\Bigg[1- e^{-i(\sum_{i=1}^3 k_i+|\vec{k}_1+\vec{k}_2+\vec{k}_3|)(\eta'-\eta_0)}\Bigg]}{16 k_1 k_2 k_3 |\vec{k}_1+\vec{k}_2+\vec{k}_3| \big[ \sum_{i=1}^3 k_i+|\vec{k}_1+\vec{k}_2+\vec{k}_3|\big]} .\label{Wofi4ds}\ee Integrating in $\eta'$ and taking $\eta -\eta_0 \rightarrow \infty$ we find
 \be C_0(\eta) = \mathcal{Z}_0\,e^{-i\Delta E_0 (\eta-\eta_0)} \label{C0fi4ds} \ee where $\mathcal{Z}_0 \sim 1-z_0$ and $\Delta E_0$ are the same as the  Minkowski spacetime results of eqs.(\ref{deltaE0fi4},\ref{z0fi4}).

As expected due to the Weyl invariance of the system, the vacuum is stable. There is only an $\eta$ dependent \emph{phase} and an overall wave function renormalization just as in Minkowski spacetime. However, the main point of this exercise is to clarify that a transition matrix element $\langle \kappa | H_I |0\rangle \neq 0$ is \emph{not} a signal of any instability of the vacuum (or for that matter of particle states). Rather, it points to the ``dressing'' of the particular state via its coupling to a continuum of states and the ensuing renormalization. Only the time dependence of the amplitude can distinguish between a true decay of the state or a simple phase and this is exactly what the  Wigner-Weisskopf approach focuses on.

\subsection{\label{subsec:cubicdeS} Vacuum amplitude and $1$-particle decay in $\varphi^3$}

A more interesting situation appears when we consider an interaction that is \emph{not} Weyl invariant, such as a cubic self-interaction $\varphi^3$. The reason for this is that as seen in eq.(\ref{rescalagds}), the interaction vertex for $\chi$ now depends on the scale factor, and hence on the conformal time $\eta$.

We follow the approach we took in the flat space case, first concentrating on the vacuum amplitude. In the Markovian approximation, the vacuum persistence amplitude satisfies
 \be \dot{C}_0(\eta)+\int^\eta_{\eta_0} \Sigma_0(\eta,\eta')\,C_0(\eta')\, d\eta' =0. \label{vacdseq}\ee The matrix elements for $\Sigma_0$ are depicted in figure \ref{fig3:fi3vaccum} but with all the lines now corresponding to the same massless, conformally coupled field $\varphi$. We find
 \be \Sigma_0(\eta,\eta') =    \frac{\lambda^2\,V}{H^2\,\eta \, \eta'}  \int \frac{d^3p}{(2\pi)^3}\int \frac{d^3k }{(2\pi)^3} \,\frac{e^{-i(p+k+|\vec{k}+\vec{p}|)(\eta-\eta')}}  {2 p\,2k \,2|\vec{k}+\vec{p}|}.  \label{sig0fi3ds}\ee The momentum integrals are similar to those appearing in the Minkowski case with massless fields and yield
 \be \Sigma_0(\eta,\eta') =     \frac{i\, \lambda^2\,V}{(4\pi)^4\,H^2\,\eta \, \eta'} \frac{1}{\Big[\eta-\eta'-i\epsilon \Big]^3}. \label{sig0fi3exds}\ee Here $\epsilon \rightarrow 0^+$ provides a ultraviolet convergence factor for the momentum integrals just as in Minkowski spacetime. We obtain
 \be W_0(\eta,\eta) =- \frac{\lambda^2\,V}{(4\pi)^4\,H^2\,\eta^4}\Bigg\{i\Big[\frac{\eta^2}{2\epsilon^2}- \ln\Big(\frac{-\eta}{\epsilon} \Big) \Big]+\frac{\eta}{\epsilon}+\frac{\pi}{2}  \Bigg\}+\cdots\,, \label{W0vacds}\ee where the dots stand for  subleading terms as $\eta,\epsilon  \rightarrow 0,\ \eta_0 \rightarrow -\infty$.

 Keeping the leading order terms in the asymptotic limits $\eta \rightarrow 0^-\,,\,\eta_0 \rightarrow -\infty\,,\,\epsilon \rightarrow 0^+$ we find
 \be C_0(\eta) \simeq e^{-i\Delta_0(\eta)} \,e^{-z_0(\eta)}, \label{asyC0ds}\ee where
 \be \Delta_0(\eta) = -\frac{\lambda^2\,V}{2(4\pi)^4\,H^2(-\eta)\,\epsilon^2}\Bigg[1-\frac{2}{3}\frac{\epsilon^2}{\eta^2}\ln\Big(\frac{-\eta}{\epsilon} \Big) \Bigg] ,\label{Delta0ds}\ee
\be z_0(\eta) = \frac{
 \lambda^2\,V}{ 2(4\pi)^4\,H^2\eta^2\,\epsilon}\,\Bigg[1+\frac{\pi\,\epsilon}{3\,\eta} \Bigg] .\label{z0ds}\ee

 The $1/\epsilon^2$ and $1/\epsilon$ ultraviolet divergences are the same as in Minkowski spacetime, when we match $t\rightarrow \infty$ to $\eta\rightarrow 0^-$. However, there are some important differences with the flat space result.  The wave function renormalization, featuring the $1/\epsilon$ divergence, is now \emph{time dependent} and grows as $\eta \rightarrow 0^-$. This is a consequence of the extra factors $1/\eta$ in the interaction vertices.   The phase $\Delta_0(\eta)$ also features a novel logarithmic ultraviolet divergence. The time evolution of the wave function renormalization function $z_0(\eta)$ is such as to force the \emph{decay} of the vacuum state and this decay is hastened as $\eta \rightarrow 0^-$, revealing the infrared nature of instability of the vacuum.

 Comparing the wave function renormalization given by eqn. (\ref{z0ds}) with the result in Minkowski space-time given by eqn. (\ref{fasefi3eps}) it is clear that the instability has nothing to do with the intrinsic instability of the $\phi^3$ theory from being unbounded below but is a direct manifestation of the cosmological expansion and breaking of conformal invariance, since the conformally invariant $\phi^4$ features an ultraviolet divergent but $\eta$ independent wave-function renormalization \emph{constant}.

The leading order ultraviolet divergence of $z_0(\eta)$ acquires an interesting interpretation if we \emph{assume} that the short distance/time ultraviolet cutoff $\epsilon$ is constant in \emph{physical coordinates}:
 \be \widetilde{\epsilon} = \frac{\epsilon}{(-H\eta)} = \mathrm{constant}. \label{physcutoff}\ee Under this assumption, the leading order term in $z_0$ becomes
  \be z_0(\eta) = \frac{
 \lambda^2\,V_{\rm phys}(\eta)}{ 2(4\pi)^4 \,\widetilde{\epsilon}}\label{minklike}\ee where
 \be V_{\rm phys}(\eta) = \frac{V}{(-H\eta)^3} \label{Vphys}\ee is the \emph{physical volume}. This is a striking result: the vacuum wave function renormalization with this regularization is the same as in Minkowski spacetime, eq.(\ref{fasefi3eps}), but in terms of the \emph{physical volume} which grows as a function of time. Therefore with this choice of regularization, the decay of the vacuum state is a consequence of the fact that the spatial volume increases with time, hence the amplitude for finding the bare state in the dressed state falls off exponentially with the physical volume.

This behavior of the vacuum state leads to some rather interesting consequences for various correlation functions that might be associated with physical observables, such as non-gaussianity. We do understand that inflaton fluctuations $\zeta$ behave as massless minimally coupled fields with derivative and non-local interactions, and so our calculations will have limited utility in understanding what happens with $\zeta$ correlators. However, we expect that the WW method can be brought to bear on this question and so we present here a calculation of the effects of the decay of the vacuum on the non-gaussian expectation value $$ \langle 0|:\chi^3(\vec{x},\eta):|0\rangle .$$

Passing to the interaction picture, $\chi(\vec{x},\eta)$ is given by eq.(\ref{quantumfieldds}) and $|0(\eta)\rangle_I = \sum_n  C_n(\eta) |n\rangle$ where the coefficients $C_n(\eta)$ are the solutions to the Weisskopf-Wigner equations (\ref{quantumfieldds},\ref{intdiffds}) with $A=0$. In an obvious notation $:\chi^3(\vec{x},\eta): \simeq (a^\dagger)^3+ (a)^3+ (a^\dagger)^2 a+ a^\dagger a^2$ and since to lowest order $H_I$ connects the vacuum to the three particle state $|\kappa\rangle = |1_{\vec{k}_1};1_{\vec{k}_2};1_{\vec{k}_3}\rangle$, (see figure \ref{fig3:fi3vaccum}), the terms that contribute are of the form $$\langle1_{\vec{k}_1};1_{\vec{k}_2};1_{\vec{k}_3}| (a^\dagger)^3|0\rangle \propto C^*_\kappa(\eta) C_0(\eta),\quad \langle 0|(a^3)| 1_{\vec{k}_1};1_{\vec{k}_2};1_{\vec{k}_3}\rangle \propto C_\kappa(\eta) C^*_0(\eta).$$ The coefficients can be computed using
 \be \langle1_{\vec{k}_1};1_{\vec{k}_2};1_{\vec{k}_3}| H_I(\eta) |0\rangle  = \frac{\lambda}{H\,\eta} \frac{(2\pi)^3\,\delta^{(3)}\big(\vec{k}_1+\vec{k}_2+\vec{k}_3\big)}{[8V^3k_1k_2k_3]^\frac{1}{2}} \,e^{i(k_1+k_2+k_3)\eta} \label{mtxele2}\ee which leads to
 \be C_{\kappa}(\eta) = -i \frac{\lambda}{H\,\eta} \frac{(2\pi)^3\,\delta^{(3)}\big(\vec{k}_1+\vec{k}_2+\vec{k}_3\big)}{[8V^3k_1k_2k_3]^\frac{1}{2}} \,\int^{\eta}_{\eta_0}\,e^{i(k_1+k_2+k_3)\eta'}\,C_0(\eta')\,\frac{d\eta'}{\eta'}. \label{C3ng}\ee Summing all contributions we finally find
 \be \langle 0|:\chi^3(\vec{x},\eta):|0\rangle = -i\frac{\lambda}{H} \int \frac{d^3k_1\,d^3k_2}{(2\pi)^6\,8 k_1k_2|\vec{k}_1+\vec{k}_2|}\,C^*_0(\eta)\int^{\eta}_{\eta_0}\,e^{i(k_1+k_2+k_3)(\eta-\eta')}\,C_0(\eta')
 \,\frac{d\eta'}{\eta'}\,+ \mathrm{h.c.}\,. \label{chi3expng}\ee The lowest perturbative order would replace $C_0 \rightarrow 1$ leading to an infrared logarithmic divergence as $\eta \rightarrow 0$. However the Weisskopf-Wigner resummation replaces $C_0$ by the solution (\ref{asyC0ds}) thereby suggesting that vacuum decay severely damps out the non-gaussian correlations.

 While this is a very preliminary result, it does suggest that the resummation provided by the Weisskopf-Wigner method may lead to important modifications of non-gaussian correlations, a topic of much importance\cite{seery,bran,giddins} that will be explored elsewhere.

Now let's turn to the behavior of the one-particle states here. Recall that in the Minkowski case, the Markovian approximation to the WW solutions allowed us to extract a decay rate for the state, as well as an energy shift and a wavefunction renormalization. The situation in de Sitter is, as might be expected by now, both more complicated and far more interesting.

For single particle states $|1_{\vec{p}} \rangle $  there are two contributions to $\Sigma(\eta,\eta')$ just as in the single particle case in Minkowski spacetime: $\Sigma(\eta,\eta')= \Sigma_0(\eta,\eta')+\Sigma_1(p,\eta,\eta')$. The disconnected vacuum contribution $\Sigma_0(\eta,\eta')$ is given by eq.(\ref{sig0fi3exds}) whereas for the connected contribution we find
 \be \Sigma_1(p,\eta,\eta') = \frac{\lambda^2}{H^2\,\eta\,\eta'}\,\int \frac{d^3k}{(2\pi)^3}\,\frac{e^{-i(k+|\vec{p}-\vec{k}|-p)(\eta-\eta'-i\epsilon)}}{2p\,2k\,2|\vec{p}-\vec{k}|} \label{sigma1dsfi31p}\ee where $\epsilon \rightarrow 0^+$ is a convergence factor that regulates the ultraviolet behavior of the  momentum integral. This integral can now be carried out leading to
 \be \Sigma_1(p,\eta,\eta') =   \frac{-i\,\lambda^2}{32\pi^2\,p\,H^2\,\eta\,\eta'\,(\eta-\eta'-i\epsilon)}, \label{sig1finds1p}\ee from which we obtain the connected contribution to $W_0$:
  \be W_{0,1}(p,\eta,\eta) = \int_{\eta_0}^{\eta}\Sigma_1(p,\eta,\eta')d\eta' =  \frac{i\,\lambda^2}{32\pi^2\,p\,H^2\,\eta^2}  \, \Bigg\{-i\,\frac{\pi}{2}-\ln\Big(\frac{-\eta}{\epsilon} \Big)- \ln\Big(\frac{\eta-\eta_0-i\epsilon}{-\eta_0} \Big) \Bigg\}. \label{Wodsfi31p}
  \ee The $\ln(\epsilon)$ reflects a logarithmic ultraviolet divergence, also seen in the one-loop self-energy, which can be renormalized by a \emph{mass renormalization}. Including a mass counterterm in the
 Lagrangian $\delta M^2 :\phi^2: /2$, yields an extra term in the interaction Hamiltonian. Upon passing to conformal time and rescaling the fields by the scale factor we have
  \be H^{ct}_I(\eta) = \frac{1}{2 H^2 \,\eta^2} \int d^3x \, \delta M^2 :\chi^2: .\label{massct}\ee This yields the matrix element
  \be -i \langle 1_{\vec{p}}|H^{ct}_I(\eta)|1_{\vec{p}}\rangle = -i\frac{\delta M^2}{2 p\,H^2\,\eta^2}. \label{mtxHct}\ee Taking this counterterm to first order and choosing
  \be \delta M^2 = \frac{\lambda^2}{16\pi^2} \, \ln\Big(\frac{\epsilon}{\mu}\Big) \label{deltam^2}\ee where $\mu$ is a renormalization scale, results in renormalizing $W_{0,1}$ by replacing $\epsilon \rightarrow \mu$ in eq.(\ref{Wodsfi31p}).

After mass renormalization we find the leading asymptotic behavior as $\eta \rightarrow 0^-,\ \eta_0 \rightarrow -\infty$
  \be C_{1_{\vec{p}}}(\eta) = C_0(\eta)\,\widetilde{C}_{1_{\vec{p}}}(\eta)\ , \quad \widetilde{C}_{1_{\vec{p}}}(\eta)= e^{-i\Delta_1(p,\eta)}\,e^{-\gamma(p,\eta)} \label{C1p}\ee
where $C_0(\eta)$ is given by the vacuum (disconnected) contribution eq.(\ref{asyC0ds}) with eqs.(\ref{Delta0ds},\ref{z0ds}) and $\widetilde{C}_{1_{\vec{p}}}(\eta)$ is the one particle connected amplitude with
\be \Delta_1(p,\eta) = -\frac{\lambda^2\,\ln\big(\frac{-\eta}{\mu}\big)}{32\pi^2\,p\,H^2(-\eta)} \label{delta1pds}\ee
\be \gamma(p,\eta) = \frac{\lambda^2}{64\pi\,p\,H^2\,(-\eta)}\,. \label{gamma1pds}\ee

The decay law $e^{-\gamma(p,\eta)}$ with eq.(\ref{gamma1pds})  can be understood without invoking the Markovian
approximation as follows: $\Sigma_1(p,\eta,\eta')$ given by eq.(\ref{sig1finds1p}) can be written as
 \be \Sigma_1(p,\eta,\eta') =   \frac{-i\,\lambda^2}{32\pi^2\,p\,H^2\,\eta\,\eta'}\,\mathcal{P}\frac{1}{(\eta-\eta')}+ \frac{ \lambda^2}{32\pi \,p\,H^2\,\eta^2}\,\delta(\eta-\eta'). \label{sig1finds1psplit}\ee The imaginary part leads to an overall $\eta$-dependent phase for $C_1$, while the real part leads to
 \be \dot{C}_1(\eta) +  \frac{ \lambda^2}{64\pi \,p\,H^2\,\eta^2}\,C_1(\eta)=0 \Rightarrow C_1(\eta) = (\mathrm{phase})\times e^{-\gamma(p,\eta)} \label{exactdecayds}\ee The extra factor $1/2$ arises because the argument of $\delta(\eta-\eta')$ vanishes at the upper limit of integration.\footnote{This can be seen by writing the integral as $\int^0_{\eta_0}\Theta(\eta-\eta')\delta(\eta-\eta')\,C_1(\eta')\,d\eta'$ with $\Theta(0)=1/2$.} Thus the asymptotic decay law is valid beyond the Markovian approximation. An as already
 pointed out in ref.\cite{boyprem,boyan} the decay of single particle states into pairs of their own quanta is a consequence of the lack of kinematic thresholds in the expanding cosmology.

 The single particle decay $1 \rightarrow 2$ in the \emph{self-interacting massless} $\phi^3$ theory in de Sitter space time is again in striking contrast with the situation in Minkowski where kinematics forbids this process. Therefore this single particle instability is again a consequence of the cosmological expansion and the manifest lack of kinematic thresholds and not in any manner associated with the intrinsic instability of
 $\phi^3$ mentioned above.

\subsection{\label{subsec:pp} Pair Production}

Just as in Minkowski spacetime, the decay of the single particle state entails the production of \emph{correlated} pairs
of particles with momenta $\vec{k},\ \vec{p}-\vec{k}$. The amplitude for pair production is obtained from the
general renormalized form eq.(\ref{renockapafin}) (see Appendix B for details) where $|A\rangle = |1_{\vec{p}}\rangle$,
\be \widetilde{C} (\vec{k} ,\vec{q} ,\eta) =    -\frac{i \,\lambda\,(2\pi)^3\,\delta^{(3)}(\vec{k} +\vec{q} -\vec{p})}{H\,\Big[8\,V^3\,p\,k \,q\Big]^\frac{1}{2}} \, \int^\eta_{\eta_0}\,e^{ i(k  +q -p)\,\eta'} ~ \widetilde{C}_{1_{\vec{p}}}(\eta') \,\frac{d\eta'}{\eta'} .\label{pairds}  \ee
The $\eta'$ integral in eq.(\ref{pairds}) cannot be done exactly; however, for weak coupling\footnote{In the $\lambda \phi^3$ theory $\lambda$ has dimensions of mass and the perturbative expansion is in terms of the dimensionless ratio $\lambda/H$.}, $\lambda^2/H^2 \ll 1$ we can provide a reliable estimate of the amplitude.

In the integrand in eq.(\ref{pairds})
\be \frac{\widetilde{C}_1(\eta)}{-\eta} = (\mathrm{phase})\times \frac{e^{-\frac{g}{(-\eta)}} }{(-\eta)},\ g = \frac{\lambda^2}{64\pi\,p\,H^2}.\label{rasca}\ee This function is $\propto 1/(-\eta)$ for $-\eta \gtrsim g $, reaches a maximum at $ -\eta \simeq  g $ and then falls off sharply to zero for  $   -\eta < g  $. This means that the integrand is dominated by the region $-\eta > g$ in which we can approximate $\widetilde{C}_{1_p} \sim 1$.

Approximating $\widetilde{C}_{1_p} \sim 1$  for $-\eta > g$ in eq.(\ref{pairds}) and assuming that all
wave-vectors are deep inside the horizon at the initial time $\eta_0$, namely $-(k +q-p)\,\eta_0 \gg 1$ we find
  \be \widetilde{C} (\vec{k} ,\vec{q};\eta) \simeq   i\frac{  \,\lambda\,(2\pi)^3\,\delta^{(3)}(\vec{k} +\vec{q} -\vec{p})}{H\,\Big[8\,V^3\,p\,k  \,q\Big]^\frac{1}{2}} \Bigg\{ -\mathrm{Ci}[- \mathrm{K} \eta]+i\,\mathrm{Si}[-\mathrm{K}\eta]-i\frac{\pi}{2} \Bigg\}, \quad \mathrm{K}= k +q-p \, \label{Cpairsasi}\ee
where $\mathrm{Ci},\mathrm{Si}$ are the cosine and sine integrals respectively. It is clear that the amplitude for pair production is largest when $K$ becomes superhorizon, namely $-K\eta \ll 1$, in which case taking $-\eta \simeq g$, we find
  \be\widetilde{C} (\vec{k} ,\vec{q};\eta)\simeq   i\frac{  \,\lambda\,(2\pi)^3\,\delta^{(3)}(\vec{k}+\vec{q}-\vec{p})}{H\,\Big[8\,V^3\,p\,k \,q\Big]^\frac{1}{2}} \Bigg\{ \ln\Big[ \frac{1}{Kg} \Big]-i\frac{\pi}{2} \Bigg\},\quad  Kg \ll 1.\label{C2large}\ee For modes that remain
 well inside the horizon during this time scale, the oscillatory phase leads to strong dephasing and cancellation of the integral since $\mathrm{Ci}[-K\eta] \rightarrow 0,\ \mathrm{Si}[-K\eta]\rightarrow \pi/2$ for $-K\eta \gg 1$.

We see then that the decay of single particle states leads to the production of particles with \emph{superhorizon} wavelengths. The total probability can be estimated by summing over the superhorizon wave-vectors that yield
the largest contribution, namely those for which $K  \lesssim 1/g$. Since in perturbation theory $g\,p =\lambda^2/64\pi H^2 \ll 1,$ it follows that these wave-vectors correspond to $k  \lesssim 1/g$. An approximate estimate for the total probability of pair production is obtained by neglecting the logarithm in eq.(\ref{C2large}), leading to
  \be \sum_{\vec{k},\vec{q}} |\widetilde{C} (\vec{k} ,\vec{q};\eta \sim 1/g) |^2 \simeq \frac{\lambda^2}{32\,\pi^2 p^2 H^2} \int^{\frac{1}{g}}_0 \Big[p+k-|p-k|\Big] dk \label{aproxpair}\ee for $g\,p \ll 1.$ The integral in eq.(\ref{aproxpair}) $\sim 2p/g$ leading to the approximate result
\be \sum_{\vec{k},\vec{q}} |\widetilde{C} (\vec{k} ,\vec{q};\eta \sim 1/g) |^2 \simeq \frac{2}{\pi} \sim 1 \label{unitds}\ee This result is a confirmation of unitarity.

It is important to establish a difference with the case of single particle decay in Minkowski spacetime analyzed in section (\ref{subsec:cubic}). Particle decay in Minkowski spacetime leads to pairs with a non-perturbatively large amplitude $\propto 1/\Gamma$ but in a narrow band of wave-vectors with a width $\propto \Gamma$. Unitarity is fulfilled by populating a narrow band of nearly resonant wave-vectors with large amplitudes. In de Sitter space-time the situation is very different; the massless particle of momentum $\vec{p}$ decays into particles with wave-vectors $\vec{k},\vec{p}-\vec{k}$ which are \emph{non-resonant}. The production amplitude begins to grow logarithmically when $\vec{k}$ crosses the horizon with $p$ either sub or super horizon. The analysis above demonstrates that particles with superhorizon wavelengths are produced with perturbatively small amplitudes, but at long time more and more particles become superhorizon and these states all become populated up to wave-vectors of order $1/g$. The sum total of probabilities is very nearly one, up to the logarithmic corrections that were neglected in the calculation above.

The detailed process of particle production actually depends on the time scale in a subtle manner.
If the wave-vector $p$ is deep \emph{inside} the Hubble radius during some early time scale, namely $-p\eta \gg 1$, the condition for the phase to be slowly varying and thus for the integral to grow logarithmically, $|(k+|\vec{p}-\vec{k}|-p)\eta| \ll 1$, implies $p \gg k$ and $|k\eta| \ll 1$, since for this configuration
$$ k+|\vec{p}-\vec{k}|-p  \simeq k(1-\widehat{\mathbf{k}}\cdot\widehat{\mathbf{p}}). $$
 This means that the decay process corresponds to the emission of  soft superhorizon quanta and a recoil of the particle, just like \emph{bremsstrahlung} as found in ref.\cite{boyan} for the decay of sub-horizon modes. After $p$ becomes superhorizon, pair production becomes more efficient by the decay into nearly back-to-back superhorizon \emph{entangled} pairs of momenta $p \ll k \leq 1/g$. Therefore particle production hastens when $p$ crosses the horizon and this decay process leads to correlated pairs of superhorizon particles.

 From the definition of $g$ in (\ref{rasca}) it follows that $p\,g = \lambda^2/64\pi H^2 \ll 1 $ for weak coupling. Therefore, in the weak coupling regime the wave-vector $p$ exits the Hubble radius well before the time scale $-\eta \sim g$ at which $\widetilde{C}_{1_p} $ begins to decrease and production of superhorizon pairs occurs during a large interval in time during which the amplitude $\widetilde{C}_{1_p} \simeq 1 $.

We conclude with the observation that at long times decay of single particles in de Sitter spacetime leads to an \emph{entangled} state of superhorizon pairs up to momentum $\sim 1/g$,
\be |1_{\vec{p}}\rangle \rightarrow \sum^{1/g}_{\vec{k}}  \widetilde{C} (\vec{k} ,\vec{p}-\vec{k};\eta \sim 1/g) ~ | 1_{\vec{k}},1_{\vec{p}-\vec{k}}\rangle \label{entangledstate}\ee where the pair amplitude is given by eq.(\ref{C2large}). Since $p\ll 1/g$ and the integral in eq.(\ref{aproxpair}) is dominated by momenta near the upper limit $\sim 1/g$, it follows that the processes with the largest amplitudes correspond to the emission of nearly  back-to-back superhorizon pairs with momenta $k \gg p$.

\section{Non-perturbative self-consistent screening?}\label{sec:conjecture}

The above results were obtained under the assumption that $W_1(t,t)$ in eq.(\ref{solumarkov}) is perturbatively small and can be neglected. While it is formally true that  $W_1(t,t)\propto \lambda^2$, the results for
the vacuum eqs.(\ref{asyC0ds}-\ref{z0ds}) might give us pause as to the validity of this argument. The volume, cutoff and
$\eta$ dependence which becomes singular at long time $\eta \rightarrow 0^-$ all conspire against the
validity of perturbation theory. However, if $z_0$ does not depend singularly on $\eta $ in the long time limit and reaches a constant, as is the case in Minkowski space time then this is simply interpreted as
a (large) renormalization and ``dressing'' of the bare state, which is renormalized by wave function renormalization as is the case in the adiabatic hypothesis. However, the strong $\eta$ dependence of $z_0(\eta)$ in (\ref{z0ds}) clearly indicates that the fully ``dressed'' vacuum state becomes orthogonal to the bare vacuum state at long time. Thus unlike the case of Minkowski space time where $W_1$ approaches a constant and its contribution can be simply lumped together in with the constant (albeit cutoff and volume dependent) wave function renormalization, in de Sitter space time $W_1(\eta,\eta)$ \emph{grows} singularly as $\eta \rightarrow 0^-$. Thus it is \emph{possible} that the denominator in eq.(\ref{asyC0ds}) compensates for the growth of the numerator \emph{slowing down} the decay of the vacuum state. To assess this possibility we now calculate $W_1$ in eq.(\ref{solumarkov}):
\be W_1(\eta,\eta)= \int^\eta_{\eta_0} W_0(\eta,\eta')d\eta' \simeq - \frac{G}{(-\eta)^3}\Bigg\{i\ln\Big( \frac{-\eta}{\epsilon\,e}\Big) - \frac{\eta}{2\epsilon}-\frac{\pi}{2}   \Bigg\},\quad G= \frac{\lambda^2\,V}{(4\pi)^4\,H^2 },\label{W1vacds}\ee where again we have neglected subleading terms. As in Appendix B, we allow for a counterterm $H_{ct}(\eta)$ in the Lagrangian to cancel the (infinite) vacuum phase, so that the equation for $C_0$ now becomes
\be \dot{C}_0(\eta) \big[1-W_1(\eta,\eta)\big]+\big[i   H_{ct}(\eta) +W_0(\eta,\eta)\big]C_0(\eta) = 0   \label{newwwds}\ee with $W_0(\eta,\eta)$ given by eq.(\ref{W0vacds}). Requiring that $H_{ct}$ cancel the divergent imaginary part of $W_0(\eta,\eta)$, we write
\be iH_{ct}(\eta)- i \frac{G}{(-\eta)^3}\Bigg[\frac{(-\eta)}{2\epsilon}-\frac{\ln(-\eta/\epsilon)}{(-\eta)} \Bigg]  \equiv i \frac{G}{(-\eta)^3}\,\Phi(\eta), \label{fieta}\ee \emph{defining} the variable $\Phi(\eta)$. Remarkably, a suitable choice of $\Phi(\eta)$ leads also to a \emph{cancelation of the real part} of $W_0$,  namely the contribution that leads to vacuum decay. With the choice of counterterm eq.(\ref{fieta}) we find
\be  \frac{iH_{ct}+W_0(\eta,\eta)}{1-W_1(\eta,\eta)} =    \frac{i\Phi(\eta)-\frac{1}{\epsilon}-\frac{\pi}{2\eta}}{\frac{(-\eta)^3}{G}+\alpha(\eta)+i\beta(\eta)}
\label{ratio}\ee where
\bea \alpha & = & - \frac{\eta}{2\epsilon}-\frac{\pi}{2}  \label{alfads}\\
\beta & = &  \ln\Big( \frac{-\eta}{\epsilon\,e}\Big)\,.\label{betads}\eea In the limit $G/(-\eta)^3 \gg 1$ the ratio in eq.(\ref{ratio}) becomes
\be \frac{i\Bigg[\Phi(\eta)\alpha(\eta)+\beta(\eta)\Big(\frac{1}{\epsilon}-\frac{\pi}{2\eta}\Big) \Bigg] +\Bigg[\Phi(\eta)\beta(\eta)-\alpha(\eta)\Big(\frac{1}{\epsilon}-\frac{\pi}{2\eta}\Big)  \Bigg] }{\alpha^2(\eta)+\beta^2(\eta)}. \label{finrat}\ee The real part of the above expression leads to the time dependent vacuum wave-function renormalization and the \emph{decay} of the vacuum state. However, it is noteworthy that the particular choice
\be \Phi(\eta) = \frac{\alpha(\eta)}{\beta(\eta)}\Big(\frac{1}{\epsilon}-\frac{\pi}{2\eta}\Big) \label{fichoice}\ee cancels the \emph{real part} leading to
\be  \frac{iH_{ct}+W_0(\eta,\eta)}{1-W_1(\eta,\eta)} = i \frac{\Big(\frac{1}{\epsilon}-\frac{\pi}{2\eta}\Big)}{\ln\Big( \frac{-\eta}{\epsilon\,e}\Big)}
\label{ratiofin}\ee namely a \emph{purely imaginary} contribution. With this choice of $\Phi$ the solution of eq.(\ref{newwwds}) becomes
\be C_0(\eta) = e^{-i\Omega(\eta) }, \quad \Omega(\eta) \simeq \frac{\eta}{\epsilon\ln\Big( \frac{-\eta}{\epsilon\,e}\Big)} + \mathcal{O}(\ln(\ln(-\eta)))\label{phaseC0fin}\ee

We interpret this result as a \emph{self-consistent} screening of the vacuum evolution. The result in eq.(\ref{ratio}) is strongly reminiscent of the behavior of a running coupling towards a fixed point; in $d=4-\epsilon$ (Euclidean) dimensions the dimensionless renormalized dimensionless coupling of a $\lambda \phi^4$ theory approaches an infrared fixed point, namely
\be g(\kappa) = \frac{\lambda \kappa^{-\epsilon}}{1+ \frac{\lambda \kappa^{-\epsilon}~I}{\epsilon}} ~~~\overrightarrow{\kappa \rightarrow 0} ~~~\frac{\epsilon}{I} \ee where $I$ is a constant and $\kappa$ an infrared scale. This suggests our conjecture: that non-perturbative self-consistent effects may lead to a fixed point in Hilbert space in which the vacuum state is stable, evolving in time only with a phase.
In this analysis we only focused on the terms that diverge  as $\eta \rightarrow 0^-,\  \epsilon \rightarrow 0$ and neglected terms that remain finite in these limits which result in a volume dependent but finite contribution to the phase and wave function renormalization which remain perturbatively small in these limits.

At this point we can only conjecture this behavior based on the above analysis, and clearly much more needs to be understood before concluding that this effect is generic to all theories. Furthermore, it could well be that higher order effects that may have been missed   could   change this picture. An important aspect that will be discussed elsewhere is the manifestation of unitarity from the contribution of states that are connected to the vacuum by the interaction.    Clearly, this phenomenon is noteworthy of further investigation and its impact on the asymptotic behavior of correlation functions require further and deeper study.

\section{Conclusions and further questions}

In this article we introduced a non-perturbative formulation to study the time evolution of quantum states in cosmological space time. The method is a quantum field theoretical generalization of the Weisskopf-Wigner\cite{ww} theory of atomic line widths which leads to a resummation of the perturbative series for the real time dynamics. In a Markovian approximation to the equations for the amplitudes, we established the correspondence between this method and Dyson-type resummations in Minkowski spacetime as well as with the dynamical renormalization group approach to real time dynamics\cite{drg}. In Minkowski spacetime this method allowed us to establish the asymptotic long time limit of the amplitudes of quantum states thereby revealing how unitarity is manifest both in the case of non-decaying vacuum and decaying single particle states. Having established the reliability of the method in Minkowski space-time, we generalized it to study the time evolution of quantum states in cosmological space time.

We then began a program to study long time and infrared divergences by implementing the non-perturbative method in two prototype quantum field theories in de Sitter space time: conformally coupled massless $\phi^4,\ \phi^3$. These are the simplest  scalar field theories that still possess many of the relevant features that are of current intense interest. The $\phi^4$ theory is conformally invariant and the asymptotic dynamics of quantum states is similar to that of Minkowski spacetime: the vacuum is ``dressed'' but stable with an amplitude having a divergent but time independent wave function renormalization and a phase which is also divergent and reflects vacuum fluctuations.

The $\phi^3$ case is more interesting; even when conformally coupled to gravity and massless, it is not a conformal invariant theory and particle production takes place. We found that the vacuum amplitude features not only a divergent phase, but also a wave function renormalization which depends singularly on (conformal) time and is proportional to the volume. This behavior entails an instability of the vacuum state which is enhanced by the volume factor. We find that in a particular regularization scheme, this instability can be simply understood from the vacuum wave function renormalization in Minkowski spacetime, but in terms of the expanding physical volume thereby arguing that the decay of the vacuum is a consequence (in this renormalization scheme) of both the wave function renormalization of the vacuum state being proportional to the spatial volume as well as the cosmological expansion of the physical volume. The potential impact of vacuum decay on a simple non-gaussian correlation function was   discussed.  We showed that the vacuum diagrams in the time evolution of \emph{excited} quantum states can be canceled systematically \emph{in perturbation theory} by renormalizing the amplitudes with a time dependent vacuum wave function renormalization. In this theory we also showed that single particle states decay into two quanta of the same field, as a consequence of the lack of kinematic thresholds, resulting in particle production. This process is hastened when the momentum of the decaying particle becomes superhorizon leading to an \emph{entangled state} of superhorizon pairs. We also analyzed how the amplitude of quantum states of these
produced pairs manifestly satisfies unitarity.

The singular time dependence of the vacuum wave function renormalization, combined with the volume and cutoff dependence all suggest a breakdown of the leading Markovian approximation. We considered the general expression eq.(\ref{solumarkov})  for the  Markovian approximation and found the remarkable result that a particular choice of counterterm in the Hamiltonian   leads to a complete  cancelation of the secular and singular contributions to  the vacuum wave function renormalization at long time and the asymptotic stability of the vacuum state. We conjecture that this is a self-consistent ``screening'' mechanism akin to the approach to a fixed point of a running coupling, in this case describing a dynamical fixed point in Hilbert space in the evolution of the state.

This body of results lead us to ask further questions that will be addressed in forthcoming work:

\begin{itemize}
\item{How to systematically calculate correlation functions in the quantum states whose amplitudes are solutions of the generalized Weisskopf-Wigner equations. This translation between the dynamics of states and correlation functions will allow a direct comparison with the dynamical renormalization group resummation program implemented at the level of correlation functions in refs.\cite{boyan,holman}.}

\item{How to consistenly go beyond the   Markovian approximation studied here, and in particular, how
to understand whether going beyond the Markovian approximation introduces important corrections to the results obtained here.}

\item{In minimally coupled massless theories, there are additional infrared divergences. How do these manifest themselves in the resummed amplitudes for the quantum states? Is there an infrared fixed point that emerges from the resummation program?}

\item{Ref.\cite{marolf} provides an intriguing, all orders result stating that the vacuum of a massive
scalar field in de Sitter space time is both stable and free of infrared singularities, and that correlation functions may be computed in a Euclidean compactification and analytically continued to Lorentzian signature without instabilities or infrared divergences. These results seemingly contradict  those of ref.\cite{polyakov}. Clearly it is important to understand whether the asymptotic screening mechanism found in our work can explain the results of ref.\cite{marolf} within the real time dynamics of states, or at least provide a bridge towards a translation between the Euclidean and real time formulation at asymptotically large time.}

\end{itemize}

Work on these and other questions is in progress and will be reported elsewhere.

\acknowledgments D.~B.is  supported by NSF grant award   PHY-0852497,  R.~H. is supported by the DOE through Grant No. DE-FG03-91-ER40682. R.~H. would also like to thank David Seery, Toni Riotto and Ramy Brustein for valuable discussions as well as the University of Sussex astrophysics group, Jerome Martin at the IAP in Paris and the CERN theory group for hospitality while this work was in progress.

\appendix
\section{\label{appsec:DRG} The DRG and the Markovian approximation}

The DRG approach also offers
a systematic method to study the evolution of quantum states\cite{drg} and provides an
alternative to the adiabatic switching-on procedure to generate
\emph{exact} eigenstates of the interacting Hamiltonian from the
eigenstates of the non-interacting system. Along the way this DRG
formulation extracts the energy shifts, wave function
renormalization constants and offers a clear description of the decay dynamics of quantum states and
its description in terms of Fermi's Golden rule.

Consider the time evolution of a quantum state in an interacting
theory with Hamiltonian
\begin{equation}\label{totalHam}
H=H_0+ \lambda \; H_I \; ,
\end{equation}
\noindent namely
\begin{equation}
|\psi(t)\rangle = e^{-iH(t-t_0)} ~|\psi(t_0)\rangle=e^{-iH_0 t} \;
U(t,t_0) \; e^{iH_0 t_0}|\psi(t_0)\rangle \; ,
\end{equation}
where we have introduced the unitary time evolution operator in
the interaction picture
\begin{equation}
U(t,t_0)= 1 -i \; \lambda \; \int^t_{t_0}H_I(t') \; dt'-\lambda^2
\; \int^t_{t_0}\int^{t'}_{t_0}H_I(t') \; H_I(t'') \; dt' \; dt'' +
{\cal O}(\lambda^3) ~~,~~ U(t_0,t_0)=1 \; , \label{unitop}
\end{equation}
\noindent with
\begin{equation}
H_I(t)= e^{iH_0 t} \; H_I  \; e^{-iH_0 t} \; .
\end{equation}
It is convenient to pass to the interaction picture in which the
states are given by
\begin{equation}
|\psi(t)\rangle_i= e^{iH_0t} \; |\psi(t)\rangle \; ,
\end{equation}
\noindent and their time evolution is given by
\begin{equation}
|\psi(t)\rangle_i= U(t,t_0) \; |\psi(t_0)\rangle_i \; .
\end{equation}

Consider
the \emph{adiabatic} time evolution operator $U_{\epsilon}(t,t_0)$ obtained
by replacing
\begin{equation}
H_I(t) \rightarrow e^{-\epsilon |t|} \; H_I(t) \; ,
\end{equation}
\noindent in the time evolution operator $U(t,t_0)$ in
eq.(\ref{unitop}). The adiabatic, or Gell-Mann-Low theorem asserts that the states
\begin{equation}
|\Psi_n\rangle = U_{\epsilon}(0,-\infty) \; |n\rangle \; ,
\end{equation}
\noindent constructed out of the eigenstates $|n\rangle$ of the
\emph{non-interacting Hamiltonian}  $H_0$ with energy $E_n$, are
\emph{exact} eigenstates of the total interacting Hamiltonian $H$,
and the exact eigenvalues are then given by\cite{gellmannlow}
\begin{equation}
E_n+\Delta E_n =
\frac{\langle\Psi_n|H|\Psi_n\rangle}{\langle\Psi_n|\Psi_n\rangle}
\; .
\end{equation}
The dynamical renormalization group provides an
illuminating alternative to this procedure. To make this approach
more clear and to establish contact with familiar results, we
now consider the case in which $|\psi(t_0)\rangle_i$ is an
eigenstate of the non-interacting Hamiltonian $H_0$, namely
\begin{equation}
|\psi(t_0)\rangle_i = |n\rangle ,\quad H_0 |n\rangle = E_n
|n\rangle \; .
\end{equation}
To highlight the main ideas of the DRG, we will focus on the
evolution of the (persistence) amplitude,
\begin{equation}
C_n(t)= \langle n|\psi(t)\rangle_i  \; ,
\end{equation}
\noindent though the time evolution of the off-diagonal overlap coefficients
$C_m(t)= \langle m|\psi(t)\rangle_i; m\neq n$ can be studied along
the same lines. In this section we restore the diagonal matrix elements of the perturbation allowing for
$\langle n|H_I|n\rangle \neq 0$ to highlight the consistency of the DRG resummation.

Inserting the identity $\sum_m |m\rangle \langle m|=1$
appropriately, we find
\begin{eqnarray}\label{Cnoft}
&& C_n(t) =  \langle n|U(t,t_0) \psi(t_0)\rangle_i =
C_n(t_0)\Bigg[1 - i\;
    \lambda \; (t-t_0) \;  \langle n|H_I|n\rangle
 -\frac{\lambda^2}{2} \; (t-t_0)^2 \;   \langle n|H_I|n\rangle^2
 \nonumber \\&& -\lambda^2 \; \int^t_{t_0} \int^{t'}_{t_0}\sum_{m\neq
n} |\langle n|H_I|m \rangle|^2 \;  e^{i(E_n-E_m)(t'-t'')} \; dt'
\; dt'' + {\cal O}(\lambda^3)\Bigg]  \; ,
\end{eqnarray}
\noindent where we have written $C_n(t_0)=\langle
n|\psi(t_0)\rangle_i$ despite our choice of initial state for
which $C_n(t_0)=1$, to emphasize that the initial amplitude
factors out.

Introducing the spectral density
\begin{equation}
\rho_n(\omega) =\sum_{m\neq n} |\langle n|H_I|m \rangle|^2
 \; \delta(\omega-E_m) \label{specQM}\;
\end{equation}
as in the main text, the straightforward
integration over the time variables $t',t''$ in eq.(\ref{Cnoft}) can be performed, yielding

\begin{eqnarray}\label{Cnoft2}
&& C_n(t) =  C_n(t_0)\Bigg\{1 - i \; \lambda \; (t-t_0) \; \langle
  n|H_I|n\rangle
 -\frac{\lambda^2}{2} \; (t-t_0)^2 \;  \langle n|H_I|n\rangle^2
 \\&& -i\, \lambda^2 \, (t-t_0) \; \int d\omega \;  \frac{\rho_n(\omega)}{(E_n-\omega)} \;
  \Bigg[1
  -\frac{\sin[(E_n-\omega)(t-t_0)]}{(E_n-\omega)(t-t_0)} \Bigg]  \\ && -\lambda^2 \,\int d\omega \;  \frac{\rho_n(\omega)}{(E_n-\omega)^2}
  \Bigg[1-\cos[(E_n-\omega)(t-t_0)]\Bigg]+{\cal O}(\lambda^3)
 \Bigg\}\; .\nonumber
\end{eqnarray}
Using eqs.(\ref{realpartofE},\ref{imagE}), we find the asymptotic long time limit,
\begin{eqnarray}\label{Cnoftt}
 C_n(t) = && C_n(t_0)\Bigg\{1 - i\; \lambda \; (t-t_0) \;  \langle
 n|H_I|n\rangle
 -\frac{\lambda^2}{2} \; (t-t_0)^2 \;  \langle n|H_I|n\rangle^2 \nonumber \\&&
 -i(t-t_0) \; \lambda^2 \; {\sum_m}'~ \frac{|\langle n|H_I|m
 \rangle|^2}{E_n-E_m} -\lambda^2 \; \pi \;  (t-t_0) \;
 \rho_n(E_n)\nonumber \\ && -\lambda^2 \; {\sum_m}'~ \frac{|\langle n|H_I|m
 \rangle|^2}{(E_n-E_m)^2}+ \mathcal{O} \left( \frac{1}{t-t_0},\lambda^3
 \right)\Bigg\} \; . \end{eqnarray}
\noindent where ${\sum_m}'$ refers to the sum over all the states
with $E_m \neq E_n$. We write the expression above as,
\begin{equation}
C_n(t)=C_n(t_0)\left[1+\lambda \; \mathcal{S}_n^{(1)}(t)+\lambda^2
\; \mathcal{S}_n^{(2)}(t)-\lambda^2 \;
\mathcal{Z}_n^{(2)}+\mathcal{O}\left( \frac{1}{t-t_0},\lambda^3
\right)\right] \label{Cnoftsec}\; ,
\end{equation}
\noindent where $\lambda^{(i)}\mathcal{S}_i^{(i)}(t)$ are secular
terms (linear and quadratic in time respectively) while the  term
$-\lambda^2 \mathcal{Z}_n^{(2)}$ in eq.(\ref{Cnoftsec}) is the
time independent term in eq.(\ref{Cnoftt}).

To use the dynamical renormalization group, we renormalize
the amplitude as
\begin{equation}
C_n(t_0)= C_n(\tau) \; R_n(\tau), ~~ R_n(\tau)= 1+ \lambda \;
r_n^{(1)}(\tau)+\lambda^2 \;  r_n^{(2)}(\tau)+{\cal O}(\lambda^3)
\; . \label{renoamp}
\end{equation}
Hence,
\begin{eqnarray}\label{cnrenor}
C_n(t)=C_n(\tau)&&\Bigg\{1+\lambda
\left[r_n^{(1)}(\tau)+\mathcal{S}_n^{(1)}(t)\right]+\lambda^2
\left[r_n^{(2)}(\tau)+ \mathcal{S}_n^{(2)}(t)+r_n^{(1)}(\tau) \;
\mathcal{S}_n^{(1)}(t)\right] \nonumber \\&&-\lambda^2 \;
\mathcal{Z}_n^{(2)}+ {\cal
  O}(\lambda^3) \Bigg\}
\end{eqnarray}
The counterterms $r_n^{(i)}(\tau)$ are chosen so as to cancel the
secular terms at a time scale $t=\tau$, leading to
\begin{eqnarray}\label{relas}
r_n^{(1)}(\tau)& = & -\mathcal{S}_n^{(1)}(\tau)\quad , \quad
r_n^{(2)}(\tau) = -\mathcal{S}_n^{(2)}(\tau)+[r_n^{(1)}(\tau)]^2\;
.
\end{eqnarray}
The independence of the amplitude $C_n(t)$ on the arbitrary time
scale $\tau$, namely $dC_n(t)/d\tau =0$ leads to the dynamical
renormalization group equation
\begin{eqnarray}
&&\dot{C}_n(\tau)\Bigg[1+\lambda \; \big( r_n^{(1)}(\tau)+
\mathcal{S}_n^{(1)}(t)\big)+{\cal O}(\lambda^3)\Bigg]+   \nonumber \\
&& C_n(\tau)\Bigg[\lambda \dot{r}_n^{(1)}(\tau)+\lambda^2 \;
\big(\dot{r}_n^{(2)}(\tau)+ \dot{r}_n^{(1)}(\tau) \;
\mathcal{S}_n^{(1)}(t) \big) +{\cal O}(\lambda^3) \Bigg]=0
\end{eqnarray}
\noindent where the dot stands for derivative with respect to
$\tau$. Keeping terms up to second order in $\lambda$ yields,
$$
\frac{\dot{C}_n(\tau)}{C_n(\tau)} = - \frac{d}{d\tau}\left\{
\lambda \; r_n^{(1)}(\tau) + \lambda^2 \; \left[ {r}_n^{(2)}(\tau)
-\frac12 \;
  \left( r_n^{(1)}(\tau) \right)^2 \right] \right \}+{\cal O}(\lambda^3)
$$
Notice that the $t$-dependent pieces cancelled out identically.
Such cancellation is necessary for the consistency of the DRG to
second order in $\lambda$ . We find using the counterterms given
by eq.(\ref{relas}), \be\label{solC} C_n(\tau) = C_n(t_0) \;
e^{\lambda \; {S}_n^{(1)}(\tau) + \lambda^2
  \;\left\{ \mathcal{S}_n^{(2)}(\tau) - \frac12 \; \left[
    {S}_n^{(1)}(\tau) \right] \right \} } \; .
\ee Using now the explicit form of the secular terms which are
read off from eq.(\ref{Cnoftt}), we finally find the DRG solution
to second order to be given by
\begin{equation}\label{solQS}
C_n(\tau)= C_n(0) \;  e^{-i \Delta E_n \tau} \; e^{-\frac{\Gamma_n
    \tau}{2}} \; ,
\end{equation}
\noindent with the energy shift and width up to
$\mathcal{O}(\lambda^2)$ given by
\begin{eqnarray}
\Delta E_n & = & \lambda \; \langle n|H_I| n\rangle + \lambda^2 \;
{\sum_m}'~ \frac{|\langle n|H_I|m
 \rangle|^2}{E_n-E_m} \label{energyshift2} \quad , \quad
 \Gamma_n =  2 \, \pi \; \rho_n(E_n)  \; .
 \end{eqnarray}
Notice that the terms in $\tau^2$ cancel identically  in the
exponent of eq.(\ref{solC}) and (\ref{solQS}). Again, such
cancellation is necessary for the consistency of the DRG\cite{drg}.

Finally, inserting eq.(\ref{solQS}) into  eq.(\ref{cnrenor}) and
choosing the arbitrary renormalization scale $\tau$ to coincide
with the time $t$, we find the asymptotic long time behavior of
the dynamical renormalization group resummed amplitude as
\begin{equation}
C_n(t) \buildrel{t \to \infty}\over= \mathcal{Z}_n \;  e^{-i
\Delta
  E_n t} \;  e^{-\frac{\Gamma_n t}{2}} \; .
\end{equation}
The energy shift $\Delta E_n$ is clearly the same as obtained in
familiar perturbation theory, the decay rate $\Gamma_n$ coincides
with the transition probability per unit time obtained from
Fermi's Golden rule, and the wave function renormalization
\begin{equation}
\mathcal{Z}_n = 1-\lambda^2 \;{\sum_m}'~ \frac{|\langle n|H_I|m
 \rangle|^2}{(E_n-E_m)^2} = \frac{\partial}{\partial
 E_n}\left(E_n+\Delta E_n\right)\label{ZQS} \; ,
 \end{equation}
is the same as obtained in usual perturbation theory and
determines the overlap between the unperturbed state and the
\emph{exact} eigenstate of the Hamiltonian. Thus, we
see that the dynamical renormalization group leads to an
alternative formulation of the construction of the exact
eigenstates which allows to extract the energy shifts, widths and
weights and that coincides with the familiar setting of quantum
mechanics. Its results also coincide with those obtained in the Markovian approximation to the WW equations considered in the main text.

This exercise has an important byproduct. By considering
the perturbative expansion up to second order in the interaction
Hamiltonian, we found secular terms that contain the square of the
first order secular terms. These must cancel consistently and
systematically in the final DRG equation since they are accounted
for in the resummation furnished by the DRG. Indeed we found such
cancellation in eqs.(\ref{solC},\ref{solQS}). This is the
equivalent of the renormalizability of the perturbative expansion.
Thus, this example not only provides a pedagogical framework to
explore and confirm the dynamical renormalization group, but also
manifestly shows the renormalizability in the sense that higher
order terms that appear in the perturbative expansion, which are
associated with the expansion of first order terms in the
solution, cancel systematically in the dynamical renormalization
group equation.

\section{Perturbative cancelation of  vacuum diagrams}

We saw in the Minkowski case that the $C_n$'s could be renormalized in a way so as to absorb the effect of disconnected vacuum diagrams. In that situation this was simple to do since the wavefunction renormalizations were constant in time. However, this does not hold in a cosmological setting. Thus, the question arises as to how to deal with the effects of vacuum diagrams. In this subsection, we give a prescription for handling this issue.

First, add the following counterterm to the Hamiltonian \be H_{ct} = -\frac{d}{d\eta}\Delta_0(\eta) \label{Hctds} \ee where $\Delta_0(\eta)$ is given by eq.(\ref{Delta0ds}). Next, follow the Minkowski procedure by redefining the amplitudes as  \be \widetilde{C}_n(\eta) = \mathcal{Z}^{-1}_0(\eta) C_n(\eta) =  e^{z_0(\eta)} C_n(\eta),\label{renods}\ee where $z_0(\eta)$ is given by eq.(\ref{z0ds}).

The set of equations for the renormalized amplitudes $\widetilde{C}$ now become
\bea \dot{\widetilde{C}}_\kappa(\eta) & = &  \Big[-i H_{ct}(\eta)+\dot{z}_0(\eta)\Big]\widetilde{C}_\kappa(\eta) - i\langle\kappa|H_I(\eta)|A\rangle\,\widetilde{C}_A(\eta) \label{renockapa}\\
\dot{\widetilde{C}}_A(\eta) & = &  \Big[-i H_{ct}(\eta)+\dot{z}_0(\eta)\Big]\widetilde{C}_A(\eta) - i \langle A|H_I(\eta)|\kappa\rangle\,\widetilde{C}_\kappa(\eta) \label{renocA}\eea

Since $\widetilde{C}_A \sim 1$ and $\widetilde{C}_\kappa \propto \lambda$ and because $H_{ct},\dot{z}_0 \propto \lambda^2$ it follows that  $$i\langle\kappa|H_I(\eta)|A\rangle\,\widetilde{C}_A(\eta) \propto \lambda $$ while  $$\Big[-i H_{ct}(\eta)+\dot{z}_0(\eta)\Big]\widetilde{C}_\kappa(\eta) \propto \lambda^3.$$ Therefore the latter terms in eq.(\ref{renockapa}) are perturbatively small, so that to leading order eq.(\ref{renockapa}) becomes:
\be  \dot{\widetilde{C}}_\kappa(\eta)   =    - i\langle\kappa|H_I(\eta)|A\rangle\,\widetilde{C}_A(\eta)\;.\label{renockapafin}\ee However, in (\ref{renocA}) both terms are of the same order and thus we must keep both of them for
consistency. Following the steps presented in the previous sections (after neglecting
the first term in the right hand side of (\ref{renockapa})), implementing the Markovian approximation, we finally obtain the equation for the single particle amplitude:
\be \dot{\widetilde{C}}_{1_{\vec{p}}}(\eta)  +  \Big[ i H_{ct}(\eta)+\dot{z}_0(\eta)+W_{0, 0}\Big]\widetilde{C}_{1_{\vec{p}}}(\eta)+W_{0,1}\widetilde{C}_{1_{\vec{p}}}(\eta) =0,\label{factords} \ee
where $W_{0, 0}$ is the vacuum contribution and $W_{0,1}$ is the single particle contribution eq.(\ref{Wodsfi31p}). It is clear that the vacuum contribution is now canceled since
\be   H_{ct}(\eta) = -\mathrm{Im}[W_{0, 0}],\quad \dot{z}_0(\eta) = - \mathrm{Re}[W_{0, 0}]\,.\label{cancelds}\ee The  solution of eq.(\ref{factords}) with eq.(\ref{cancelds}) is given by eq.(\ref{C1p}). It should be clear that the same procedure cancels the vacuum contribution in the equation of motion for the \emph{renormalized} amplitudes for all excited states.

\end{document}